\documentclass[]{aa}
\usepackage{graphicx}
\begin{document}

\title{The interaction of young massive stars with their environment
\thanks{Based on observations with the Atacama Pathfinder EXperiment telescope (APEX, Chile) and the Swedish-ESO
Sub-millimeter
Telescope (SEST, ESO/La Silla, Chile).}
\thanks{Figure~\ref{spec} is only available in electronic form at http://www.edpsciences.org}}
\subtitle{A millimeter and submillimeter line study of NGC\,6334 FIR\,II}

\author{J.S.~Zhang\inst{1,2,3} \and  C.~Henkel \inst{1} \and  R.~Mauersberger\inst{4} \and Y.-N. Chin
\inst{5} \and K.M.~Menten \inst{1} \and A.R.~Tieftrunk \inst{6} \and A.~Belloche \inst{1}}
\offprints{J.S. Zhang, GuangZhou University, dennyzhang998@hotmail.com}

\institute{ Max-Planck-Institut f{\"u}r Radioastronomie, Auf dem H{\"u}gel 69, D-53121 Bonn, Germany 
       \and 
            Purple Mountain Observatory, Chinese Academy of Sciences, NanJing, 210008, China
       \and 
            Center for Astrophysics, GuangZhou University, GuangZhou, 510006, China 
       \and 
            Instituto de Radioastronom{\'i}a Millim{\'e}trica (IRAM), Avenida Divina Pastora 7, Local 20, E-18012 Granada, Spain 
       \and 
            Department of Physics, Tamkang University, 251-37 Tamsui, Taipeh county, Taiwan 
       \and 
            Im Acker 21b, D-56072 Koblenz, Germany}

\date{received ; accepted}

\abstract{
Using the 15-m Swedish ESO Sub-millimeter Telescope (SEST), CO, HCN, and HCO$^+$ observations of the galactic star-forming region NGC\,6334\,FIR\,II
are presented, complemented by [C\,{\sc i}] $^3$P$_1-^3$P$_{0}$ and $^3$P$_2-^3$P$_{1}$ data from the Atacama Pathfinder Experiment (APEX 12-m 
telescope). Embedded in the extended molecular cloud and associated with the H\,{\sc ii} region NGC\,6334--D, there is a molecular ``void''. 
[C\,{\sc i}] correlates well with $^{13}$CO and other molecular lines and shows no rim brightening relative to molecular cloud regions farther off 
the void. While an interpretation in terms of a highly clumped cloud morphology is possible, with photon dominated regions (PDRs) reaching deep into 
the cloud, the data do not provide any direct evidence for a close association of [C\,{\sc i}] with PDRs. Kinetic temperatures are $\sim$40--50\,K 
in the molecular cloud and $\ga$200\,K toward the void. CO and [C\,{\sc i}] excitation temperatures are similar. A comparison of molecular and 
atomic fine structure line emission with the far infrared and radio continuum as well as the distribution of 2.2$\,\mu$m H$_2$ emission indicates 
that the well-evolved H\,{\sc ii} region expands into a medium that is homogeneous on pc-scales. If the H$_2$ emission is predominantly shock 
excited, both the expanding ionization front (classified as subsonic, ``D-type'') and the associated shock front farther out (traced by H$_2$) 
can be identified, observationally confirming for the first time a classical scenario that is predicted by evolutionary models of H{\sc ii} 
regions. Integrated line intensity ratios of the observed molecules are determined, implying a mean C$^{18}$O/C$^{17}$O abundance ratio of 
4.13 $\pm$ 0.13 that reflects the $^{18}$O/$^{17}$O isotope ratio. This ratio is consistent with values determined in nearby clouds. Right 
at the edge of the void, however, the oxygen isotope ratio might be smaller.

 }

\titlerunning{Molecules and atoms in NGC\,6334 FIR\,II}
\authorrunning{J.S. Zhang et al.}

\maketitle

\section{Introduction}

Young massive stars greatly affect their environment, either impeding or triggering future star formation. Theory (e.g., Franco et al. 1990)
predicts that soon after the radiation field of a hot massive star has been turned on, its energetic photons create a supersonic (``R-type'') 
ionization front that moves through the surrounding molecular gas, leaving it hot and ionized. The speed of this front is continuously reduced by 
geometric dilution and recombination. At a time when the speed is not much larger than the speed of sound, i.e., when the ``initial'' Str{\"o}mgen 
radius is reached, the large pressure gradient across the ionization front is supposed to drive a supersonic shock wave that sweeps, accelerates, 
and compresses the surrounding medium. The ionization front further decelerates, becomes subsonic (``D-type''), and follows the shock front, while 
neutral gas accumulates in the interphase between the fronts.

Complications arise when the surrounding gas is clumpy. If the ionization or shock front encounters a strong negative density gradient,
e.g., at the edge of the parent cloud, the ionized gas may expand supersonically (``champagne phase''). Another complication arises when young massive
stars form in clusters so that scenarios based on a single dominating star may be misleading. Poorly collimated mainly low velocity stellar winds
from young massive stars may also contribute significantly to the budget of incoming energy (e.g., Tenorio-Tagle \cite{tenorio}; Churchwell
\cite{church}).

Since it is known that molecular gas is highly clumped (e.g., Wilson \& Walmsley \cite{wilsonw}), the less energetic UV radiation ($\sim$6--13.6\,eV)
emitted by the young star(s) may leak deeply into the cloud. Such ``photon-dominated regions'' (PDRs) may thus make up a large fraction of the 
volume of a star-forming cloud (e.g., Hollenbach \& Tielens \cite{hollen}). The UV radiation from the star as well as associated shock fronts heat the 
dust and the gas. Molecules will be photodissociated and surface layers of dust grains will evaporate. The particles generated in such ways are 
ready to form new molecular species characteristic of a PDR or shocked environment.

Many star-forming regions have been studied in some detail. Usually it is found that the geometry of the scrutinized region is irregular. In 
the following we present a study of an object of particularly simple geometry, the periphery of an almost circular region devoid of molecular gas. 
This region forms the eastern part of NGC\,6334\,FIR\,II, the outstanding source of far infrared emission in the prominent southern star forming 
region NGC\,6334 (Loughran et al. \cite{loughran}). Here we present maps of the region in a number of molecular transitions ($\lambda$$\sim$3 
and 1.3\,mm) that are complemented by profiles from the 492 and 809\,GHz lines of neutral carbon (C\,{\sc i}). Observational methods are described 
in Sect.\,2. New data are presented in Sect.\,3, while Sect.\,4 provides physical parameters, including isotope ratios determined from CO. 
Section\,5 summarizes the main results.

\section{Observations}

\subsection{SEST observations}

The $J=1-0$ lines of $^{12}$CO, $^{13}$CO, C$^{18}$O, C$^{17}$O, HCN, and HCO$^+$, as well as the $J=2-1$ line of $^{13}$CO have been observed 
with the 15-m Swedish-ESO Sub-millimeter Telescope (SEST, Booth et al. 1989) in May 1993 (C$^{18}$O) and May 1994. For the 3\,mm observations, 
a Schottky receiver was used. System temperatures, $T_{\rm sys}$, were typically 480\,K on a main-beam brightness temperature ($T_{\rm mb}$) 
scale. 1.3\,mm data were obtained with an SIS receiver with $T_{\rm sys}$$\sim$1600\,K.

The chopper wheel method was used for calibration. The intensity scale was converted to a $T_{\rm mb}$ scale (see Downes 1989; Rohlfs \&
Wilson 1996) assuming main-beam efficiencies of 0.75, 0.70, and 0.60 at $\nu \sim 90\,$GHz, $110-115$\,GHz, and 220\,GHz, respectively (SEST 
Handbook, 1993). The full width to half power (FWHP) beam-width was 57$''$ for HCN and HCO$^+$, 45$''$ for the $J=1-0$ lines of CO and its 
rare isotopomers, and 24$''$ for the $J=2-1$ transition of $^{13}$CO.

The pointing of the telescope was regularly determined by mapping the SiO masers toward VX\,Sgr and AH\,Sco. We estimate the pointing to be
correct within 10$''$ (but see C$^{17}$O in Sect.\,4.7), while the calibration uncertainty is estimated to be $\pm$15\%.
As backends, we employed acousto-optical spectrometers with 1000 or 1652 contiguous channels, a channel spacing of 43\,kHz, and a spectral
resolution of 80\,kHz (0.27, 0.21, and 0.11\,km\,s$^{-1}$ at 90, 112, and 220\,GHz, respectively). The $J=1-0$ transitions of $^{12}$CO and
$^{13}$CO were measured in a dual frequency switching mode, with frequency offsets of $\pm 15\,\rm MHz$. The other measurements were carried
out in a position switching mode, the reference position being at an offset of $(\Delta \alpha, \Delta \delta)=(-20',20')$. The integration
times per point were usually 15 to 60 seconds; only for C$^{17}$O were integration times longer, with typically 5 minutes per position. From
all the spectra, linear baselines have been subtracted. A summary of observational parameters is presented in Table\,\ref{obspa}. Spectra
reduced using the GILDAS software (e.g., Guilloteau \& Lucas 2000) are shown in Fig.\,\ref{spec}\footnote{Throughout the paper, B1950.0 
coordinates were used to facilitate comparisons with previously published results (e.g., Brooks \& Whiteoak \cite{brooks}; Ezoe et al. 
\cite{ezoe}; and earlier publications).} of the appendix that is available electronically.

\subsection{APEX observations}

The $^3$P$_1-^3$P$_0$ and $^3$P$_2-^3$P$_1$ fine-structure lines of neutral atomic carbon (C\,{\sc i}) were observed in April 2006, using FLASH
(First Light Apex Submillimeter Heterodyne instrument; Heyminck et al. 2006) on the 12-m APEX (Atacama Pathfinder Experiment; G{\"u}sten et al.
\cite{guesten}) telescope. FLASH is a dual-channel heterodyne instrument operating simultaneously in the 420--500\,GHz and the 780--880\,GHz
atmospheric windows. Double sideband system temperatures were $\sim$1000--2000\,K and 3800--8000\,K on a $T_{\rm mb}$ scale, respectively, mainly 
depending on elevation. The measurements were carried out in a position switching mode with the reference position at an offset of ($\Delta \alpha, 
\Delta \delta)$ = (20$'$,0$'$). On-source integration times were 30 to 90 seconds per position; the central position was observed repeatedly to check 
calibration, resulting in a total on-source integration time of about 300 seconds. Observational parameters are summarized in Table\,\ref{obspa}.

The data were calibrated to a $T_{\rm mb}$ scale adopting main-beam efficiencies of 0.60 and 0.43 and forward hemisphere efficiencies of 0.95
at $\nu \sim 492\,$GHz and 809\,GHz, respectively. Full width to half power (FWHP) beam-widths are 14\arcsec\ and 8\arcsec\ (G{\"u}sten et al. 2006). 
The absolute calibration uncertainty is estimated to be $\pm$25\%; the relative calibration uncertainty is about $\pm$15\%.

As backend a fast-fourier-transform spectrometer was used, containing 16384 channels and covering a bandwidth of 1\,GHz (Klein et al. 2006).
The resulting channel spacing is 61\,kHz (0.04 and 0.02\,km\,s$^{-1}$ at the lower and higher frequencies, respectively). Pointing and focus
sources were SgrB2(N) and Venus.

\begin{table}
\caption[]{Observed lines toward NGC6334
\label{obspa}}
\begin{flushleft}
\begin{tabular}{lrrrlr}
\hline\noalign{\smallskip}
Transition& Frequency&   Beam     & $L^{a)}$     & $\eta$$_{mb}^{b)}$&  Time$^{c)}$  \\
          & \multicolumn{1}{c}{(GHz)} & \multicolumn{1}{c}{(\arcsec)}  & \multicolumn{1}{c}{(pc)} &
          & \multicolumn{1}{c}{(s)}         \\
\hline\noalign{\smallskip}
SEST 15-m                                                                       \\
\noalign{\smallskip}
\noalign{\smallskip}
HCN$(1-0)$             & 88.632 & 57 & 0.48 & 0.75   &   30           \\
HCO$^{+}$(1$-$0)       & 89.188 & 57 & 0.48 & 0.75   &   30           \\
$^{12}$C$^{18}$O(1$-$0)&109.782 & 45 & 0.38 & 0.7    &   60           \\
$^{13}$C$^{16}$O(1$-$0)&110.201 & 45 & 0.38 & 0.7    &   15           \\
$^{12}$C$^{17}$O(1$-$0)&112.359 & 45 & 0.38 & 0.7    &  300           \\
$^{12}$C$^{16}$O(1$-$0)&115.271 & 45 & 0.38 & 0.7    &   15           \\
$^{13}$C$^{16}$O(2$-$1)&220.399 & 24 & 0.20 & 0.6    &   30           \\

\hline\noalign{\smallskip}

APEX 12-m       \\
 \noalign{\smallskip}
 \noalign{\smallskip}
$[$C\,{\sc i}$]$$^{3}$P$_{1}$$-$$^{3}$P$_{0}$&492.160&14& 0.11 & 0.60   &   30           \\
$[$C\,{\sc i}$]$$^{3}$P$_{2}$$-$$^{3}$P$_{1}$&809.341& 8& 0.07 & 0.43   &   30           \\
\hline
\noalign{\smallskip}
\noalign{\smallskip}
\end{tabular}
\end{flushleft}

a) To estimate the linear size of the respective beam, $D$=1.7\,kpc is adopted (Neckel \cite{neckel}). \\
b) Main beam efficiencies; for references see Sect.\,2. \\
c) On-source integration times: For [C\,{\sc i}], integration times are typically 30 seconds per scan, but the central position has been observed
several times (see Sect.\,2.2). \\

\end{table}

\section{Results}

\subsection{Molecular line emission}

\begin{figure*}
\resizebox{16cm}{21cm}{\includegraphics{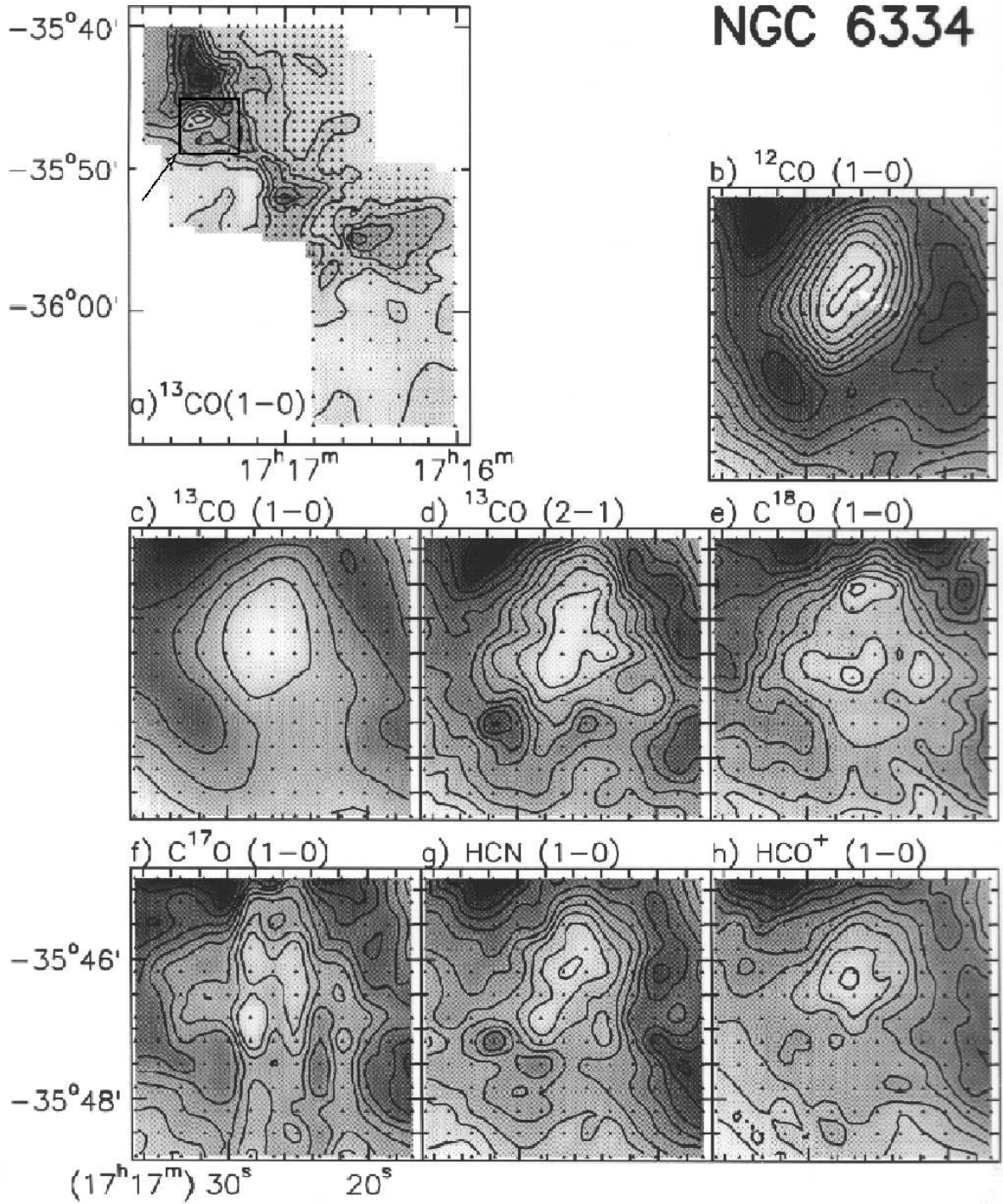}}
\caption[]{\label{cnt} Contour maps of integrated intensity, $\int T_{\rm MB}{\rm d}v$. 1\arcmin\ corresponds to 0.5\,pc. The measured positions
are indicated by markers. {\bf a)} Large-scale $^{13}$CO $J$=1$-$0 map. The lowest contour is $8\,\rm K\,km\,s^{-1}$, the contour increment is
$18\,\rm K\,km\,s^{-1}$, and the rms noise is $1.11\,\rm K\,km\,s^{-1}$. The arrow points toward the area shown in the other maps of this figure,
including the emission of seven molecular lines centered at $\alpha_{1950}=17^{\rm h} 17^{\rm m} 26\fs 7$, $\delta_{1950} = -35^{\rm o} 46'50''$
($\alpha_{2000}=17^{\rm h} 20^{\rm m} 47\fs 9$, $\delta_{1950} = -35^{\rm o} 49'48''$). {\bf b)} $^{12}$CO $J=1-0$
(lowest contour: $99\,\rm K\,km\,s^{-1}$, increment: $33\,\rm K\,km\,s^{-1}$, rms noise: $5.85\,\rm K\,km\,s^{-1}$). {\bf c)} $^{13}$CO $J$=1--0
(lowest contour: $8\,\rm K\,km\,s^{-1}$, increment: $18\,\rm K\,km\,s^{-1}$, rms noise: $1.82\,\rm K\,km\,s^{-1}$). {\bf d)} $^{13}$CO $J$=2--1
(lowest contour: $35\,\rm K\,km\,s^{-1}$, increment: $21\,\rm K\,km\,s^{-1}$, rms noise: $0.91\,\rm K\,km\,s^{-1}$).  {\bf e)} C$^{18}$O $J$=1--0
(lowest contour: $0.5\,\rm K\,km\,s^{-1}$, increment: $2\,\rm K\,km\,s^{-1}$, rms noise: $0.52\,\rm K\,km\,s^{-1}$). {\bf f)} C$^{17}$O $J$=1--0
(lowest contour: $0.1\,\rm K\,km\,s^{-1}$, increment: $0.5\,\rm K\,km\,s^{-1}$, rms noise: $0.14\,\rm K\,km\,s^{-1}$). {\bf g)} HCN $J$=1--0
(lowest contour: $14\,\rm K\,km\,s^{-1}$, increment: $7\,\rm K\,km\,s^{-1}$, rms noise: $1.04\,\rm K\,km\,s^{-1}$). {\bf h)} HCO$^+$ $J$=1--0
(lowest contour: $10\,\rm K\,km\,s^{-1}$, increment: $4.5\,\rm K\,km\,s^{-1}$, rms noise: $0.78\,\rm K\,km\,s^{-1}$).}
\end{figure*}

We mapped the large-scale distribution of the $^{13}$CO $J=1-0$ emission over a region of roughly 10$'$$\times$$30'$ (Fig.\,\ref{cnt}a) 
covering almost the entire ``molecular ridge'' of NGC\,6334 that is characterized by a position angle of about 45$^{\circ}$ (e.g., Kraemer \& 
Jackson \cite{kraemer99}).  At a distance of 1.7$(\pm0.3)$\,kpc (Neckel \cite{neckel}), 1\arcmin\ corresponds to a linear size of 0.5\,pc. The 
map spacing is $40''$ near the ridge and $2'$ further away from the ridge. The northeastern part of the map shows a small inconspicious ``void'' 
surrounded by extended molecular emission. Covering this area that is marked by a square in Fig.\,\ref{cnt}a, maps of $^{12}$CO, $^{13}$CO, 
C$^{18}$O, C$^{17}$O, HCN, and HCO$^+$ were taken on a finer grid consisting of $13\times 13$ spectra. These are shown in Figs.\,\ref{cnt}b-h. 
The spacing of $20''$ implies full sampling for all lines except $^{13}$CO $J=2-1$. In these more detailed maps, the void becomes a prominent 
feature common to all molecular lines observed by us across the eastern part of the FIR\,II region.

\begin{figure}
\resizebox{\hsize}{!}{\rotatebox[origin=br]{-90.0}{\includegraphics{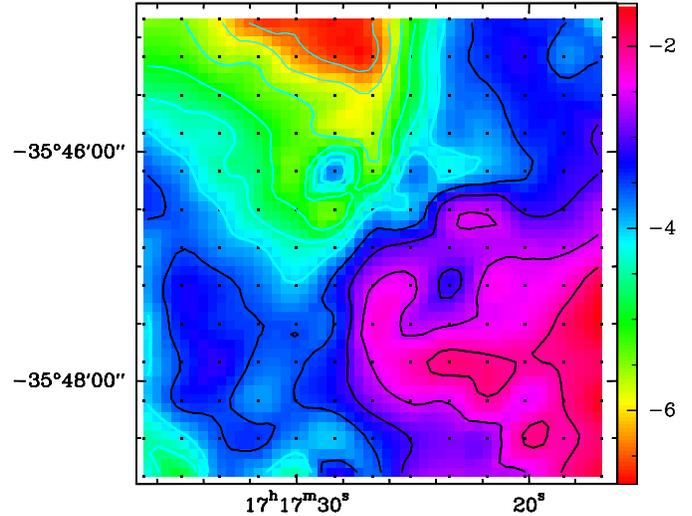}}}
\caption{Local standard of rest ($V_{\rm LSR}$) velocity distribution of $^{13}$CO(1-0) across the eastern part of NGC\,6334 FIR\,II  
(B1950.0 coordinates). Contours: --6.5 to --2.0\,km\,s$^{-1}$ by 0.5\,km\,s$^{-1}$. 1\arcmin\ corresponds to 0.5\,pc.
\label{13co10v}}
\end{figure}

Line intensity contrasts are high. Ratios of integrated intensity between the position of strongest emission in the mapped area (offset relative
to the position of minimum emission: ($\Delta \alpha,\Delta \delta$) = (60\arcsec,80\arcsec)) and the minimum itself at $\alpha_{1950}$ = 17$^{\rm h}$
17$^{\rm m}$ 26\fs 7, $\delta_{1950}$ = --35$^{\circ}$ 46$'$ 10$''$ ($\alpha_{2000}$ = 17$^{\rm h}$ 20$^{\rm m}$ 47\fs 9, $\delta_{2000}$ =
--35$^{\circ}$ 49$'$ 08$''$) are 4, 10, 6, 4, 6, and 6 for CO, $^{13}$CO $J$=1--0, $^{13}$CO $J$=2--1, C$^{18}$O, HCN, and HCO$^+$, respectively
(because of its weakness, C$^{17}$O is not included here; the error in the ratios can be deduced from the 15\% calibration uncertainties given 
in Sect.\,2.1 and is therefore of order $\pm$20\%). The ratios are lower limits because not only the main beam but also the sidelobes may 
contribute to the signal seen at the center of the molecular void. The size of the region is 1$\,.\!\!^{\prime}$5$\pm$0$\,.\!\!^{\prime}$2, if 
measured where integrated intensities reach twice the minimum value. This holds for CO, for the rare CO isotopomers, as well as for the high density 
tracers HCN and HCO$^+$. The size is significantly larger than the beam (Sect.\,2.1), so that estimates based on $^{13}$CO $J$=1--0 and 2--1 data 
(FWHP beamwidths: 45$''$ and 24$''$) do agree even without beam deconvolution. While in CO the center of the molecular void is surrounded by a 
closed ring of strong CO emission, the rare CO isotopomers show an opening of the ring toward the south. The high density tracers HCN and HCO$^+$ 
also show such an opening toward the south-east and east.  Here only half of the ring is prominent, even though the minimum is clearly identified. 
From this we conclude that molecular column densities are particularly high in the east, north, and west, while densities reach the highest values near 
the northern and western edges of the molecular void.

Calculating the mean velocity (first moment) of the observed emission, we show the velocity distribution of the $^{13}$CO $J$=1-0 line in
Fig.\,\ref{13co10v}. The other lines show a similar distribution. Molecular gas with ``low'' velocities dominates the northeast (down to
--6.5\,km\,s$^{-1}$), while gas with ``high'' velocities is observed in the southwest (up to --2\,km\,s$^{-1}$). The velocity gradient is approximately
5\,km\,s$^{-1}$\,pc$^{-1}$; linewidths are $\sim$5--6\,km\,s$^{-1}$.

\begin{figure}
\resizebox{8.5cm}{16cm}{\includegraphics{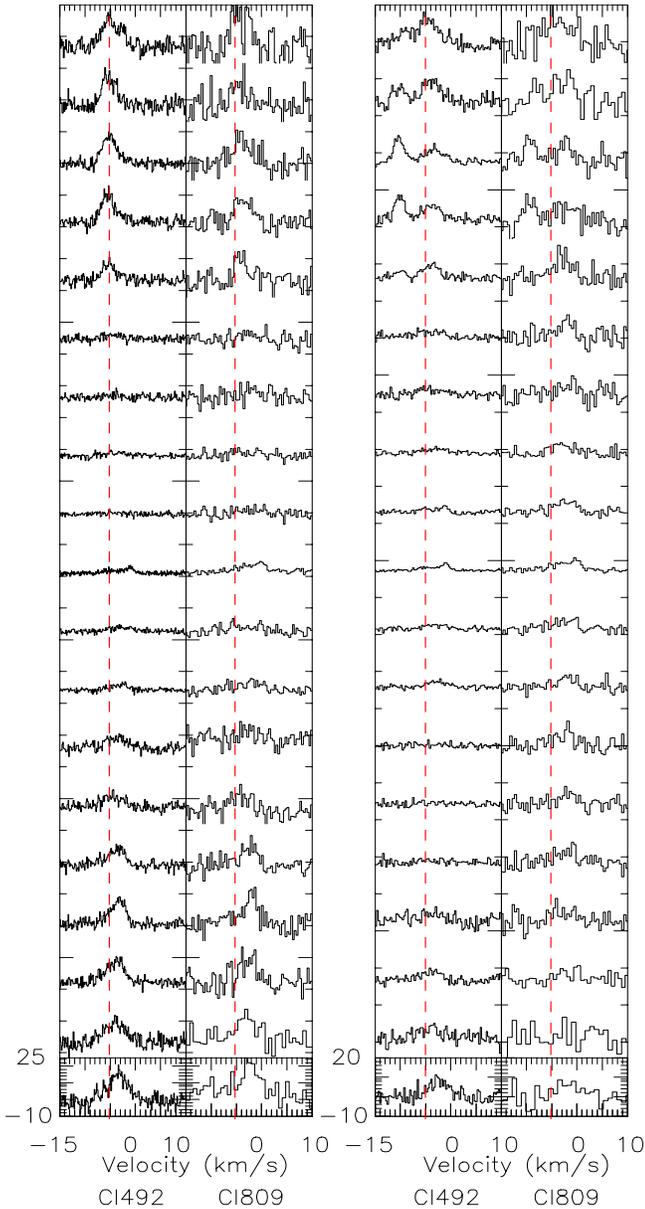}}
\caption[]{\label{CI} Spectra of the [C\,{\sc i}] $^3$P$_1-^3$P$_{0}$ 492\,GHz line (channel spacings 0.15\,km\,s$^{-1}$ (left panel) and 0.30\,km\,s$^{-1}$ 
(right panel)) and the $^3$P$_2-^3$P$_{1}$ 809\,GHz line (channel spacings 0.68\,km\,s$^{-1}$ (left panel) and 0.90\,km\,s$^{-1}$ (right panel)) 
toward the molecular void region. In the left panel, west is up and east is down; in the right one north is up and south is down. The central position is 
at $\alpha_{1950}=17^{\rm h} 17^{\rm m} 26\fs 7$, $\delta_{1950} = -35^{\rm o} 46'10''$ ($\alpha_{2000}=17^{\rm h}20^{\rm m} 47\fs 9$, $\delta_{2000}
=- 35^{\rm o} 49'08''$), the spacing is 10$''$. The ordinate is in units of main-beam brightness temperature (K). To facilitate comparisons, $V_{\rm LSR}$ 
= --5\,km\,s$^{-1}$ is marked by dashed vertical lines. Adopted rest frequencies are 492.160651 and 809.341970\,GHz.} 
\end{figure}

\subsection{Atomic carbon emission}

To study the neutral atomic component of the molecular void, we observed the two [C\,{\sc i}] lines of the $^3$P fine structure system at 492
($^3$P$_1-^3$P$_0$) and 809\,GHz ($^3$P$_2-^3$P$_1$). [C\,{\sc i}] is considered to be a classical PDR tracer (e.g., Hollenbach \& Tielens \cite{hollen}). 
The cross consisting of 19 positions along the E-W and N-S directions is shown in Fig.\,\ref{CI}. The spacing is 10\arcsec. While one might have 
expected particularly strong emission at the inner portion of the molecular shell, with weaker emission deeper inside the molecular cloud, this 
is not seen.  Instead, the two atomic carbon lines show a trend that is similar to that of the molecular emission lines. The line intensity increases 
when leaving the inner region of the void, reaching a high plateau where the surrounding molecular cloud is located.

As in the case of the rare CO isotopomers, the molecular edge toward the south exhibits weaker emission than the positions at the other
ends of the cross. Because of the much smaller beamsize of the [C\,{\sc i}] measurements when being compared with the molecular data
(Sect.\,3.1), the size of the region devoid of strong emission as well as spatial fine structure can be evaluated with higher accuracy. 
Along the E-W axis, we find an extent of 80\arcsec\ for the molecular void; along the N-S axis the corresponding value is 100\arcsec, 
defined in the same way as for the molecular emission in Sect.\,3.1. Toward the west, the edge of the void is particularly pronounced. 
Here the [C\,{\sc i}] emission rises by a factor of $>2$ within 10\arcsec\ (0.1\,pc). This is the region that appears to be characterized 
by high column densities {\it and} high particle densities (Sect.\,3.1). A significant difference between the distributions of the 492 
and 809\,GHz lines of [C\,{\sc i}] is not apparent. Within the limits of spatial resolution, the morphology of [C\,{\sc i}] is compatible 
with that of the molecular tracers.

For the central position, the main beam brightness temperatures are $\sim$4 and $\sim$6\,K for [C\,{\sc i}] $^3$P$_1-^3$P$_{0}$ and $^3$P$_2-^3$P$_1$,
respectively, while temperatures of order 20--30\,K are reached in the outermost eastern and western positions. The contrast between the molecular
edge and the central emission features is $\sim$5$\pm$1. This is consistent with the corresponding molecular ratios choosing the same reference
positions. While the mean velocities agree with those of the molecular lines (see Fig.\,\ref{13co10v}), the [C\,{\sc i}] lines are slightly
narrower, typically $\sim$4\,km\,s$^{-1}$. This is likely caused by the different beamwidths (Table~\ref{obspa}).

\section{Analysis and discussion}

In the following we give a brief overview on NGC\,6334 with an emphasis on FIR\,II. Then physical parameters of the gas near the void are
evaluated, starting with an analysis of [C\,{\sc i}] that is followed by CO and the high density tracers HCN and HCO$^+$. After searching for
traces of molecular outflows seen in CO, $^{18}$O/$^{17}$O ratios and the evolutionary stage of the entire region are also discussed.

\begin{figure*}
\resizebox{22cm}{22cm}{\includegraphics{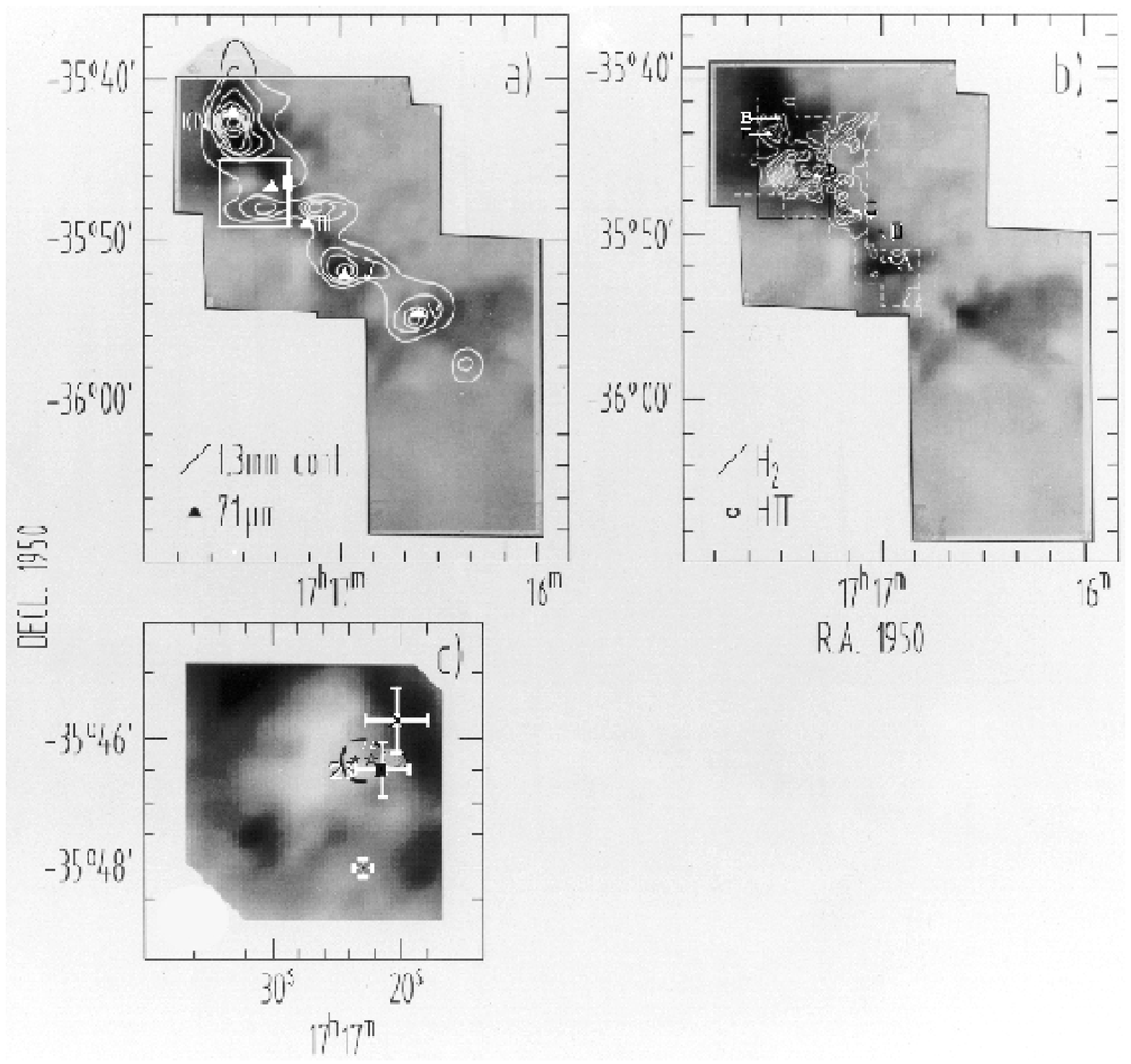}}
\vspace{-5cm}
\caption[]{\label{finder} Finder charts of the features on a larger (a, b) and smaller (c) scale toward NGC\,6334. The inset in a) (white) and
b) (black) shows the region sketched in more detail in c). The grey scale outlines the integrated $^{13}$CO $J$=1--0 intensity (this paper).
{\bf a)} Contours show the distribution of the 1\,mm continuum emission (Gezari \& Blitz 1990). Triangles with Roman numbers mark the positions
of the 71$\,\mu$m peaks (Loughran et al. 1986). {\bf b)} White contours show the distribution of shocked H$_2$ gas (Straw \& Hyland \cite{strawb}),
with regions of strongest emission being emphasized by white diagonal lines. Prominent H\,{\sc ii} regions are marked by A, B, C, D, E, F (see
Rodr\'{\i}guez et al. \cite{rodriguez}). {\bf c)} The Str{\"o}mgren sphere of the H\,{\sc ii} region D is shown as a black dashed ellipse for
$n$(H$_2$)=10$^{4}$\,cm$^{-3}$ ($R_{\rm str}$ $\propto$ $n$(H$_2$)$^{-2/3}$; see Sect.\,4.8). The two $K$-band stars IRS--23 (early B) and
IRS--24 (likely O6.5) are given as smaller and larger black asterisks, respectively (e.g., Straw et al. \cite{strawc}). The 1\,mm peak identified
by Gezari \& Blitz (\cite{gezarib}) is plotted as a small dot in the southwestern part of the map. The 42$\mu$m FIR peak (Loughran et al.
\cite{loughran}) is marked as a black square with white error bars and the H$_2$O maser identified by Moran \& Rodr\'{\i}guez (\cite{moran}) is
displayed as a cross, also with white error bars, near the northwestern edge of the map.}
\end{figure*}

\subsection{NGC\,6334: A brief overview}

NGC\,6334\footnote{The nomenclature of molecular, maser, and infrared features is quite complex (see the appendix in Kuiper et al. 1995). We
follow recent publications in using the nomenclature given by McBreen et al. (\cite{mcbreen}) and Gezari (\cite{gezari}) for the infrared
sources and Rodr\'{\i}guez et al. (1982) for the H\,{\sc ii} regions.} hosts some of the most active sites of star formation in the Galaxy.
Embedded in an extended region of H$\alpha$ emission that encompasses almost 1$^{\circ}$ (Meaburn \& White \cite{meaburn}; Burton et al.
\cite{burton}) lies a giant molecular cloud (GMC). Its ridge of length $\sim$30\arcmin\ at P.A. $\sim$ 45$^{\circ}$ (see Figs.\,\ref{cnt}a
and \ref{finder}a for $^{13}$CO $J$=1--0 emission) shows spectacular, complex structure including dense molecular clumps, maser sources,
H\,{\sc ii} regions, and luminous sources of dust emission (Fig.\,\ref{finder}). The ridge has been detected in many energy bands ranging
from low frequency radio waves up to 80\,keV (Burton et al. \cite{burton}; Caproni et al. \cite{caproni}; Kraemer et al. \cite{kraemer00};
McCutcheon et al. \cite{mccutcheon}; Persi et al. \cite{persi}; Sandell \cite{sandell}; Sarma et al. \cite{sarma}; Brooks \& Whiteoak
\cite{brooks}; Carral et al. \cite{carral}; Bykov et al.  \cite{bykov}; Ezoe et al. \cite{ezoe}; Hunter et al. \cite{hunter}; Leurini et al.
\cite{leurini}; Schilke et al. \cite{schilke}; and references therein).

Along the active ridge, NGC\,6334 FIR\,II is the outstanding source of far infrared emission ($L_{\rm FIR}$ $\sim$ 3$\times$10$^5$\,L$_{\odot}$;
see Fig.\,\ref{finder}c). Its peak shifts with increasing wavelength to the west, with the 21$\mu$m and 134$\mu$m peaks being displaced
by about 2\arcmin\ (Loughran et al. \cite{loughran}). Slightly southeast of the infrared peak lies an H\,{\sc ii} region (NGC\,6334--D) of
diameter 2\arcmin\ that is known to show an almost circular, featureless thermal radio continuum distribution (Brooks \& Whiteoak
\cite{brooks} and references therein) and that is characterized by an exceptionally low extinction (the ``molecular void''; see Straw
\& Hyland \cite{strawa} for an extinction map). This H\,{\sc ii} region appears to be powered by a highly obscured O6.5 star, possibly the
most massive star of the entire GMC (Neckel \cite{neckel}; Rodr\'{\i}guez et al.  \cite{rodriguez}; Loughran et al. \cite{loughran};
Straw et al. \cite{strawc}).

Straw \& Hyland (\cite{strawb}) imaged the H$_2$ 2.12$\mu$m $v$=1--0 S(1) line that shows an intense arc at the southern, eastern, and 
northern edges of the radio continuum source (see also Fig.\,\ref{finder}b) being part of a ring surrounding the H\,{\sc ii} region.  
In addition, they took near infrared broad band spectra toward selected positions. Analyzing line ratios they conclude that most of the 
H$_2$ emission is caused by shocks rather than by fluorescence triggered by UV radiation. However, extinction is severely affecting measured 
line ratios, so that the case is not entirely clear (see also Burton et al. \cite{burton}). The good correlation between this H$_2$ ring and 
the dust extinction led Straw \& Hyland (\cite{strawb}) to propose that young stars have swept the inner part of the ring (the molecular 
void) free of gas and dust and that the H$_2$ emission is powered by the interaction between the stellar wind and the remaining molecular 
gas. The shell is also seen in the 3.3$\mu$m emission of polycyclic aromatic hydrocarbon (PAH) (Burton et al. \cite{burton}). Overall, 
FIR\,II appears to belong to the more evolved centers of activity in NGC\,6334 because the far infrared source and the associated 
H\,{\sc ii} region are more extended than other similar targets in the region (Loughran et al. \cite{loughran}; Straw et al. 
\cite{strawc}) and because there is no indication of particularly low molecular densities in the surrounding environment (e.g., Sect. 3.1).

\subsection{Global distribution of the molecular gas}

The distribution of our large-scale $^{13}$CO $J$=1--0 map (Fig.\,\ref{cnt}a) agrees well with the CO $J$=2--1 and CS $J$=3--2
maps presented by Kraemer \& Jackson (\cite{kraemer99}). Our $^{13}$CO $J$=1--0 map shows pronounced peaks toward NGC\,6334 FIR I
(up to $\sim$150\,K\,km\,s$^{-1}$), IV and V, while FIR II and III are associated with tongues of ridge emission extending southwards
from FIR I and northeastwards from FIR IV, respectively. The less well sampled southwestern part of our $^{13}$CO map, not covered by
Kraemer \& Jackson (\cite{kraemer99}), shows no significant additional features. For the narrow ridge, our map is also consistent with 
the $^{13}$CO $J$=2--1 map of Kraemer \& Jackson (\cite{kraemer99}; their Fig.\,3) that covers a smaller region.

\begin{table*}
\caption[]{[C\,{\sc i}] excitation parameters for five key positions (center and outermost positions toward E, S, W, and N; see Fig.\,\ref{CI})$^{*}$
\label{parameters}}
\begin{flushleft}
\begin{center}
\begin{tabular}{lccccccccc}
\hline\noalign{\smallskip}

Pos. & \multicolumn{2}{c}{$\int{T_{\rm mb}\,\,{\rm d}v}$} & \multicolumn{2}{c}{$T_{\rm mb}$} & $R$ & $T_{\rm ex}$&${\tau}_{1-0}$ &
       ${\tau}_{2-1}$ & $N_{\rm CI}$        \\
     &  (1--0)        &       (2--1)    &        (1--0)           &             (2--1)       &     &             &               &
                      &                     \\
     &                        \multicolumn{4}{c}{(K)}                                        &     &     (K)     &               &
                      & (10$^{17}$cm$^{-2}$)\\
\hline\noalign{\smallskip}

(0,0)& 14.3$\pm$1.7 &  28.2$\pm$4.8  &  4.1$\pm$1.0 &  6.5$\pm$1.7 & 1.97$\pm$0.30 & $\ga$200  & $\la$0.02      & $\la$0.04       &  $\sim$2  \\
 E   & 80.9$\pm$3.9 &  79.2$\pm$13.5 & 18.6$\pm$4.7 & 27.1$\pm$6.8 & 0.98$\pm$0.15 & 51$\pm$13 & 0.6(+0.5,--0.2)& $\ga$0.8        &  $\sim$15 \\
 W   & 70.3$\pm$3.7 &  72.1$\pm$13.6 & 18.1$\pm$4.5 & 26.6$\pm$6.7 & 1.02$\pm$0.15 & 53$\pm$9  & 0.6(+0.2,--0.1)& 1.3(+2.2,--0.4) &  $\sim$13 \\
 N   & 92.4$\pm$4.6 & 113.3$\pm$13.9 & 16.5$\pm$4.5 & 21.0$\pm$5.3 & 1.23$\pm$0.18 & 72$\pm$20 & 0.3(+0.2,--0.1)& 0.5(+0.4,--0.2) &  $\sim$15 \\
 S   & 40.6$\pm$4.5 &  39.2$\pm$14.0 & 11.0$\pm$2.8 & 13.7$\pm$3.4 & 0.96$\pm$0.14 & 49$\pm$10 & 0.3(+0.2,--0.1)& 0.6(+0.3,--0.2) &  $\sim$6  \\

\noalign{\smallskip} \hline \noalign{\smallskip}
\end{tabular}
\end{center}
\end{flushleft}

$^{*}$ Columns: \\
  Column~1: Offset position; E, W, N, and S denote the outermost positions shown in Fig.\,\ref{CI}, with offsets of 90\arcsec\ along the four
                  cardinal directions. \\
  Columns~2 and 3: Integrated main beam brightness temperatures of the $^3$P$_1-^3$P$_0$ and $^3$P$_2-^3$P$_1$ lines; errors are determined from
                  Gaussian fits. \\
  Columns~4 and 5: Main beam brightness temperatures of the $^3$P$_1-^3$P$_0$ and $^3$P$_2-^3$P$_1$ lines; errors reflect uncertainties
                  in absolute calibration ($\pm$25\%; see Sect.\,2) and are not providing information on signal-to-noise ratios.  \\
  Column~6: Integrated intensity ratio ([C\,{\sc i}] $^3$P$_2-^3$P$_1$ / [C\,{\sc i}] $^3$P$_1-^3$P$_0$) with 15\% uncertainties as estimated in
                  Sect.\,2. \\
  Column~7: Excitation temperature, derived from $T_{\rm ex}$ = 38.8/ln(2.11/$R$) (Schneider et al. 2003; for $R$, see Col.\,6). \\
  Columns~8 and 9: Optical depths, derived from ${\tau}_{1-0}$ = --ln(1--$T_{\rm mb}$(1--0) $\times$ (exp(23.62/$T_{\rm ex}$)--1)/23.62)
                  and ${\tau}_{2-1}$ = --ln(1--$T_{\rm mb}$(2--1) $\times$ (exp(38.8/$T_{\rm ex}$)--1)/38.8) (Schneider et al. 2003). \\
  Column~10: [C\,{\sc i}] column density, derived from
                  $N_{\rm CI}$ = 5.94$\times$10$^{15}$ $\times$ ${\tau}_{1-0}$/(1--exp(--${\tau_{1-0}}$)) $\times$
                  [(1+3\,exp(--23.6/$T_{\rm ex}$) + 5\,exp(--62.4/$T_{\rm ex}$))/3\,exp(--23.6/$T_{\rm ex}$)] $\times$ $\int{T_{\rm mb}(1-0){\rm d}v}$
                  (see Schneider et al. 2003).
\end{table*}

\subsection{[C\,{\sc i}] excitation, optical depth, column density, and correlation with $^{13}$CO}

The C\,{\sc i} triplet with lines near 492 and 809\,GHz is a useful tracer of excitation. Our measured [C\,{\sc i}] $^3$P$_{2}-^{3}$P$_{1}$ 
brightness temperatures are slightly higher than those of the [C\,{\sc i}] $^{3}$P$_{1}$$-$$^{3}$P$_{0}$ line (see below). This is consistent 
with results from other galactic star-forming regions (e.g., Schneider et al. 2003). Obtaining Gaussian fits to the [C\,{\sc i}] spectra, we 
calculated integrated line intensities for five key positions, the central one and the four outermost positions along the four cardinal directions 
(see Fig.\,\ref{CI}). Assuming local thermodynamic equilibrium (LTE), optically thin line emission, and uniform cloud 
coverage, the integrated intensity ratios allow us to determine the excitation temperature and optical depths of both transitions as well 
as the beam averaged C\,{\sc i} column density (e.g., Stutzki et al. \cite{stutzki}; Wei{\ss} et al. \cite{weiss}). Table~\ref{parameters} 
displays the resulting parameters and the equations used. The ratios of integrated line intensities ($\sim$1.2--1.5) and excitation 
temperatures (50--70\,K) are similar for the four edge positions. Toward the central position of the molecular void, however, the line 
ratio is almost two and the excitation temperature must be $\ga$200\,K. This is in accordance with the temperature of the photodissociated 
gas assumed by Kraemer et al. (\cite{kraemer00}), 250\,K. While in view of possible systematic calibration errors the exact value given in 
Table~\ref{parameters} may be debatable, we note that the integration time toward the central position is much higher than toward all other 
positions, so that a low signal-to-noise ratio does not affect the deduced parameters. Furthermore, all adjacent positions still belonging to
the inner part of the void confirm, albeit with lower signal-to-noise ratios, the high [C\,{\sc i}] $^3$P$_2-^1$P$_{0}$/$^3$P$_1-^{3}$P$_{0}$ 
line intensity ratio. Since the [C\,{\sc i}] lines are likely thermalized ($T_{\rm ex}$ $\sim$ $T_{\rm kin}$), we conclude that the inner part 
of the studied region is dominated by gas with temperatures that are much higher than those in the predominantly molecular outer regions.  
While the [C\,{\sc i}] optical depth at the center of the void is one to two orders of magnitude below those at the molecular edges, the total 
C\,{\sc i} column density is as a consequence of the high excitation, only a factor of several smaller.

\begin{figure}
\resizebox{\hsize}{!}{\includegraphics{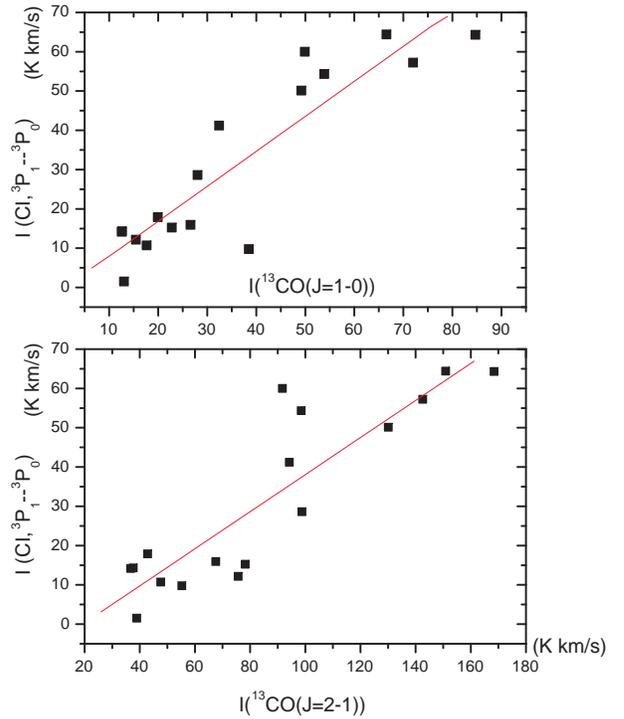}}
\caption[]{\label{CICO} A comparison of integrated intensities between the [C\,{\sc i}] $^{3}$P$_{1}$$-$$^{3}$P$_{0}$ and the $^{13}$CO $J$=1--0
and 2--1 lines.}
\end{figure}

With respect to intensity and morphology, the 492 and 809\,GHz lines of neutral carbon are known to be correlated with low-$J$ CO emission (e.g.,
Ikeda et al. \cite{ikeda}). The [C\,{\sc i}] $^3$P$_1-^3$P$_0$ and $^3$P$_2-^3$P$_1$ fine structure lines have upper energy levels of 23.6\,K and
62.5\,K above the ground state, respectively. Critical densities are of order 10$^{3}$ and 10$^{4}$\,cm$^{-3}$ for collisions with H$_{2}$. These
critical densities agree with those of the lowest rotational transitions of optically thin CO isotopomers (e.g., Mao et al. \cite{mao}). Toward
the molecular void, main beam brightness temperatures of the [C\,{\sc i}] $^{3}$P$_{1}$$-$$^{3}$P$_{0}$ line are similar to those of the
$^{13}$CO $J$=1--0 transition and about a factor of $\sim$1.6 lower than those of $^{13}$CO $J$=2--1. Correlations between $^{13}$CO and
[C\,{\sc i}] emission are shown in Fig.\,\ref{CICO}. Ignoring differences in beam size we compare the integrated intensities for the 17
positions with both $^{13}$CO and [C\,{\sc i}] emission measured. Results from linear fitting are $I({\rm CI (^{3}P_{1}-^{3}P_{0})}) =
(0.89\pm0.11)I({\rm ^{13}CO (J=1-0)})+(-1.00\pm4.67)$ and $I({\rm CI (^{3}P_{1}-^{3}P_{0})}) = (0.47\pm0.06)I({\rm ^{13}CO (J=2-1)})+(-9.12\pm6.10)$,
with correlation coefficients of $r$=0.88 and $r$=0.90, respectively. Again, this is consistent with results from other molecular clouds
(e.g., Plume et al. \cite{plume}; Ikeda et al. \cite{ikeda}).

\subsection{CO}

\begin{figure}
\resizebox{\hsize}{!}{\rotatebox[origin=br]{-89.9}{\includegraphics{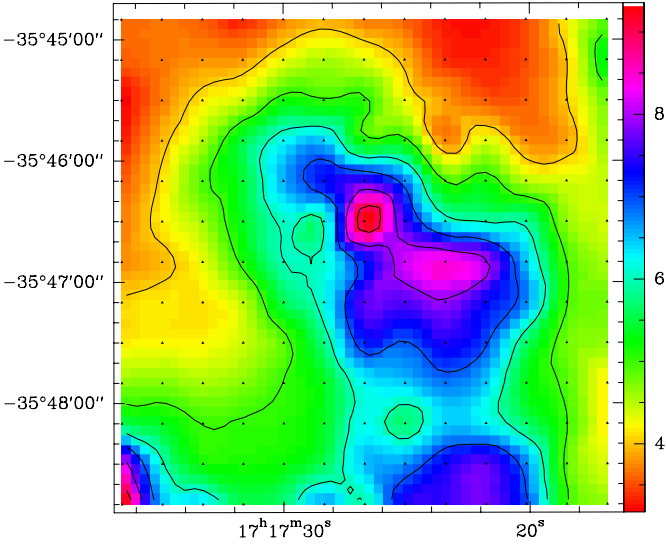}}}
\caption[]{\label{co13coratio} Integrated $^{12}$CO/$^{13}$CO $J$=1--0 line intensity ratios near N\,6334--D. Contours: 4--9, increment: 1.}
\end{figure}

\begin{figure}
\resizebox{\hsize}{!}{\rotatebox[origin=br]{-89.9}{\includegraphics{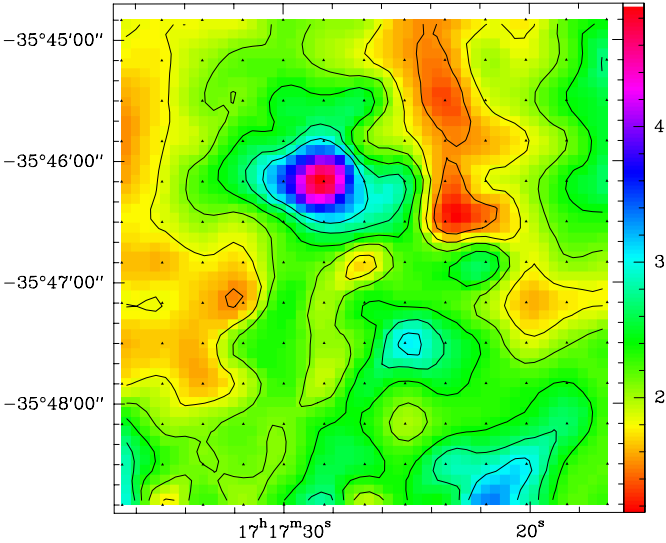}}}
\caption[]{\label{13coratio} Integrated $^{13}$CO $J$=2--1/$J$=1--0 line intensity ratios near NGC\,6334--D. Contours: 1.5--3.0, increment: 0.3.}
\end{figure}

Not only atomic but also molecular lines can be a measure of excitation conditions. Figures~\ref{co13coratio} and \ref{13coratio} show the
integrated CO/$^{13}$CO $J$=1--0 and $^{13}$CO $J$=2--1/$J$=1--0 line ratios for the area covering the molecular void. $^{12}$CO/$^{13}$CO
ratios peak, not surprisingly, at the very center of the map where the expanding H\,{\sc ii} region has removed most of the molecular gas along
the line of sight. Here the line ratios almost reach a value of 10, while values of $\sim$4 characterize the partial ring of relatively
high column density that opens up toward the south (see also Sect.\,3.1).

CO must be optically thick because CO/$^{13}$CO line intensity ratios are much smaller than the ``local'' $^{12}$C/$^{13}$C ratio of $\sim$70
(e.g., Wilson \& Rood \cite{wilsonr}). From the peak CO temperature, assuming uniform coverage of the beam, the CO excitation temperature
$T_{\rm ex}$ can then be estimated by
$$
T_{\rm mb,CO} = \frac{{\rm h}\nu}{\rm k} \left[ \left( {\rm e}^{{\rm h}\nu/{\rm k}T_{\rm ex}}-1 \right)^{-1}
                                             - \left( {\rm e}^{{\rm h}\nu/{\rm k}T_{\rm cmb}}-1 \right)^{-1} \right].
$$
With $T_{\rm mb}$ $\sim$ 40--50\,K ($T_{\rm cmb}$=2.73\,K) in the molecular cloud surrounding the void we get $T_{\rm ex}$ values that are of the
same order as $T_{\rm mb}$. CO must be thermalized because high optical depths reduce the low critical densities even below the values mentioned
in Sect.\,4.3. Being thermalized and optically thick, the measured $T_{\rm mb,CO}$ values provide an estimate of the kinetic temperature of
the molecular gas. $T_{\rm kin}$ $\sim$ 40--50\,K. In the case of self-absorption, we would have only obtained lower limits to $T_{\rm ex}$ and
$T_{\rm kin}$. The CO profiles (Fig.\,\ref{spec}) show, however, no evidence for this effect.

In the C\,{\sc i} analysis presented in Sect.\,4.3 (the E, W, N, and S positions of Table~\ref{parameters}) we obtained similar excitation temperatures
as for CO.  Does the good agreement between C\,{\sc i} and CO excitation provide a further argument that CO self-absorption can be neglected, that
C\,{\sc i} may also be thermalized, and that the APEX and SEST beams are approximately uniformly filled with C\,{\sc i} and molecular gas, respectively?

Assuming that (1) the $^{13}$CO lines are optically thin, that (2) the cloud covers the beam uniformly, and (3) neglecting the microwave background
(a good approximation for $T_{\rm ex}$$\ga$10\,K), a comparison of the $^{13}$CO $J$=2--1 and 1--0 lines provides yet another measure of excitation.
Here again line intensity ratios peak near the center of the molecular void, reaching $\sim$4, while the ratio decreases to $\sim$2 in the vicinity
of the partial ring of high column density (Fig.\,\ref{13coratio}). We then obtain with
$$
T_{\rm ex,13CO} = -10.6/\left[ {\rm ln} \left (\frac{T_{\rm mb,2-1}}{4 \times T_{\rm mb,1-0}} \right ) \right ]
$$
excitation temperatures near infinity at the center of the void (within the errors consistent with the high excitation obtained from C\,{\sc i}; see
the (0,0) position in Table~\ref{parameters}) and values $\sim$15\,K near the ``molecular ring''. These latter temperatures disagree with those obtained
from CO and C\,{\sc i}. Is this due to subthermal excitation of $^{13}$CO, is it caused by the fact that C\,{\sc i} traces mostly the UV irradiated 
surfaces of clouds, or is it an optical depth effect? If $^{13}$CO is not optically thin, our derived value for $T_{\rm ex,13CO}$ would only be a lower 
limit.

Since the critical densities of CO, $^{13}$CO, and C\,{\sc i} are similar, different excitation temperatures must either be related to different spatial
distributions or to optical depth effects. In the case of line saturation, photon trapping slows radiative rates by $\sim$$\tau^{-1}$, leading to a
greater dominance of collisional processes that tend to shift excitation temperatures upwards, toward $T_{\rm kin}$. With a CO/$^{13}$CO abundance ratio
of 70 and line intensity ratios of order 4, $\tau$(CO $J$=1--0) $\sim$20 and $\tau$($^{13}$CO $J$=1--0) $\sim$0.3. Applying an LVG code, with
H$_2$--CO collision rates from Flower (\cite{flower}) and an ortho/para H$_2$ abundance ratio of 3 (not a critical parameter), we find a density
of $n$(H$_2$) $\sim$10$^{3}$\,cm$^{-3}$ and an optical depth of order 1.5--2.0 for the $^{13}$CO $J$=2--1 line. Thus the $^{13}$CO $J$=2--1 line
is likely moderately optically thick and the equation for $T_{\rm ex,13CO}$ indeed provides too low a value. In addition, $^{13}$CO is not fully 
thermalized ($T_{\rm ex,13CO}$ $<$ $T_{\rm ex,12CO}$). 

To summarize, there is evidence that C\,{\sc i} and CO show excitation temperatures of $T_{\rm ex}$ $\sim$ 40--50\,K, which is comparable
to the kinetic temperature. $N$($^{13}$CO) $\sim$ 10$^{17}$\,cm$^{-2}$ and $n$(H$_2$) $\sim$ 10$^3$\,cm$^{-3}$. These values hold, however, only for
the parent molecular cloud. Within the molecular void, excitation and kinetic temperatures are far higher but too uncertain to derive reliable (column)
densities. According to Straw \& Hyland (\cite{strawa}; their Fig.\,9), $A_{\rm v}$ $<$10$^{\rm m}$ in the void, corresponding to a column density
of $N$(H$_2$) $<$ 3$\times$10$^{22}$\,cm$^{-2}$ (Bohlin et al. \cite{bohlin}) or $N$($^{13}$CO) $<$ 4$\times$10$^{16}$\,cm$^{-2}$ ([H$_2$]/[CO] =
10$^4$, [CO]/[$^{13}$CO] = 70). IRS--24, likely the star that is responsible for the molecular void, is reported to show an extinction of
$\ga$28$^{\rm m}$ (Straw et al. \cite{strawc}), corresponding to a $^{13}$CO column density of $N$($^{13}$CO) $\ga$10$^{17}$\,cm$^{-2}$. This is
compatible with the characteristic value already calculated for the molecular cloud. Apparently, the star is situated near the western edge of
the void where column densities are high (IRS--23, a little nearer to the void, shows $A_{\rm v}$ $\sim$ 15$^{\rm m}$; Straw et al. \cite{strawc}).
In view of the high obscuration of IRS--24, the star appears to be located behind the bulk of the molecular gas.

\subsection{Other molecules}

\begin{figure}
\resizebox{\hsize}{!}{\rotatebox[origin=br]{-89.9}{\includegraphics{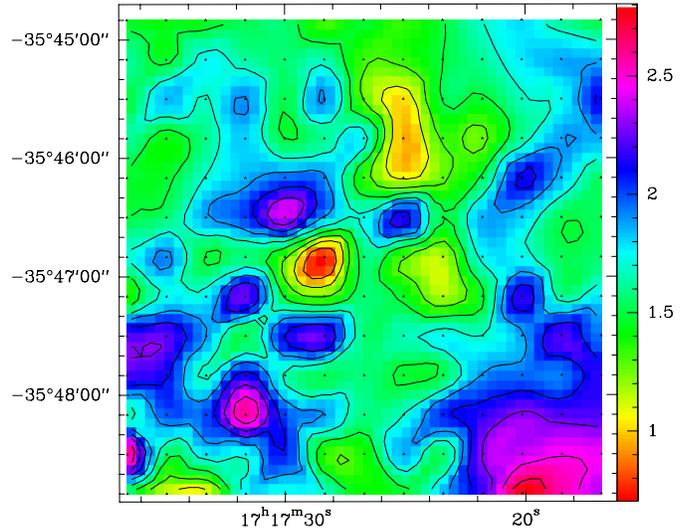}}}
\caption[]{\label{hcnhco+ratio} Integrated HCN/HCO$^+$ $J$=1--0 line intensity ratios observed with a beam size of 57\arcsec\ toward NGC\,6334--D. 
Contours: 0.7--2.5, increment: 0.2.}
\end{figure}

The excitation temperatures calculated for CO and its isotopomers cannot be used for HCN and HCO$^+$, which have much higher critical densities. An
analysis of the HCN hyperfine components shows that the HCN $J$=1--0 transition, wherever it is strong enough to show its characteristic triple,
is optically thin. Adopting $T_{\rm ex}$ = 10\,K (a reasonable guess implying subthermal excitation), characteristic column densities are of order
$N$(HCN) $\sim$ 7$\times$10$^{13}$\,cm$^{-2}$ and $N$(HCO$^+$) $\sim$2$\times$10$^{13}$\,cm$^{-2}$ for the molecular cloud. To reach $T_{\rm mb}$
values of 5\,K (or more in a few positions), LVG calculations with $T_{\rm kin}$ = 45\,K and collision rates of Sch{\"o}ier et al. (\cite{schoier})
require $n$(H$_2$) $\ga$ 10$^5$\,cm$^{-3}$. This is a much higher density than that derived from CO, indicating a high degree of small-scale clumping
(see also Sect.\,4.8). The HCN/HCO$^+$ $J$=1--0 line intensity ratio is displayed in Fig.\,\ref{hcnhco+ratio}. In the molecular cloud, HCN is stronger, 
in particular in the south-western region that forms a part of the molecular ridge. Signal-to-noise ratios are low for both lines near the void, so 
that their ratio shows a large scatter. Nevertheless, an inspection of the individual profiles confirms the impression given by Fig.\,\ref{hcnhco+ratio}, 
that HCO$^+$ becomes about as strong as HCN. This may be a consequence of relatively high kinetic temperatures, strong radiation fields, and low 
densities near the edge of the molecular cloud (e.g., Fuente et al. \cite{fuente}; Chin et al. \cite{chin}; Brouillet et al. \cite{brouillet}; 
Christopher et al. \cite{christopher}).

\begin{table}
\caption[]{Average integrated line ratios toward NGC6334--D$^{\rm a)}$
\label{ratios}}
\begin{flushleft}
\begin{tabular}{lr}
\hline\noalign{\smallskip}
      Transitions                               &        Ratio   \\
\hline\noalign{\smallskip}
$I$(CO $J$=1--0)/$I$($^{13}$CO $J$=1--0)        &  5.3$\pm$0.1   \\
$I$($^{13}$CO $J$=2--1)/$I$($^{13}$CO $J$=1--0) &  2.2$\pm$0.1   \\
$I$(CO $J$=1--0)/$I$(HCN $J$=1--0)              &  7.1$\pm$0.2   \\
$I$(HCN $J$=1--0)/$I$(HCO$^+$ $J$=1--0)         &  1.7$\pm$0.1   \\

\hline
\noalign{\smallskip}
\noalign{\smallskip}
\end{tabular}
\end{flushleft}

a) Given errors were obtained from Gaussian fits (standard deviation).  \\

\end{table}

\begin{figure}
\resizebox{\hsize}{!}{\includegraphics{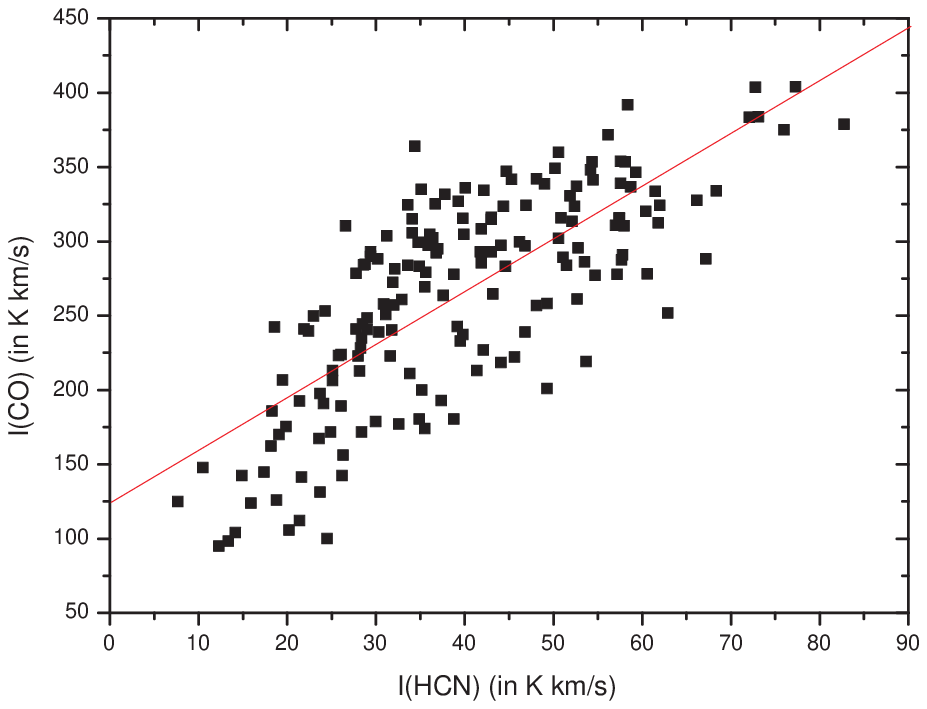}}
\resizebox{\hsize}{!}{\includegraphics{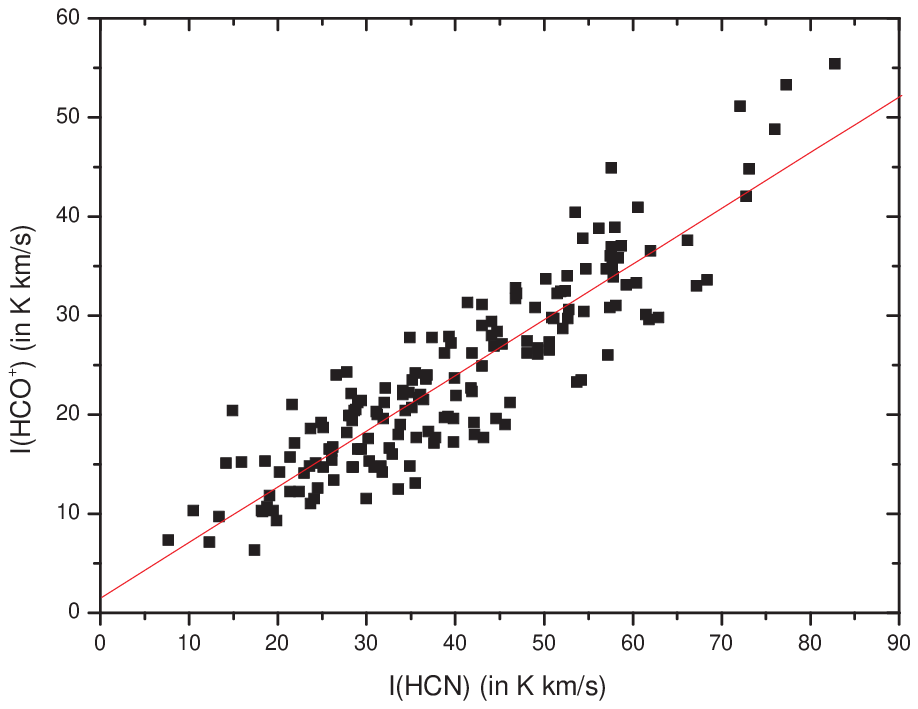}}
\caption[]{\label{ii} Comparison of integrated intensites of the $J$=1--0 transitions of CO, HCO$^{+}$, and HCN. The lines mark least square fits
with $I$(CO $J$=1--0) = (3.55$\pm$0.24) $I$(HCN) + (123$\pm$10) (correlation coefficient: 0.75) and $I$(HCO$^+$ $J$=1--0) = (0.56$\pm$0.02)
$I$(HCN) + (1.47$\pm$0.89) (correlation coefficient: 0.90).} 
\end{figure}

Averaged integrated intensity ratios are summarized in Table~\ref{ratios}. The intensity ratio of HCN to CO could be a qualitative and even a
quantitative measure of pressure (Helfer \& Blitz \cite{helfer}). Our mean values are consistent with observational results from other clouds (e.g.,
Pirogov \cite{pirogov}; Helfer \& Blitz \cite{helfer}). Unlike the HCN and HCO$^+$ $J$=1--0 lines, CO $J$=1--0 is usually opaque tracing lower
density gas. Hence a stronger correlation is expected between HCN and HCO$^+$ than between CO and HCN or HCO$^+$. This is confirmed by
Fig.\,\ref{ii}.

\subsection{Outflows in NGC\,6334 FIR\,II}

Straw et al. (\cite{strawb}) found an arc of shocked vibrationally excited H$_2$ enveloping the H\,{\sc ii} region NGC\,6334-D (Sect.\,4.1) that
coincides with the molecular void. This was interpreted in terms of molecular gas swept away by an outflow powered by the central star(s). Brooks
\& Whiteoak (\cite{brooks}) mention that so far no further observations have been made to confirm the presence of this outflow. Although we
have not tested this idea by directly observing shocked H$_2$ with high velocity resolution, we have collected well-resolved spectra from the 
CO $J$=1--0 transition, which is {\it the} outflow tracer for cool molecular gas.

\begin{figure}
\resizebox{\hsize}{!}{\rotatebox[origin=br]{-89.9}{\includegraphics{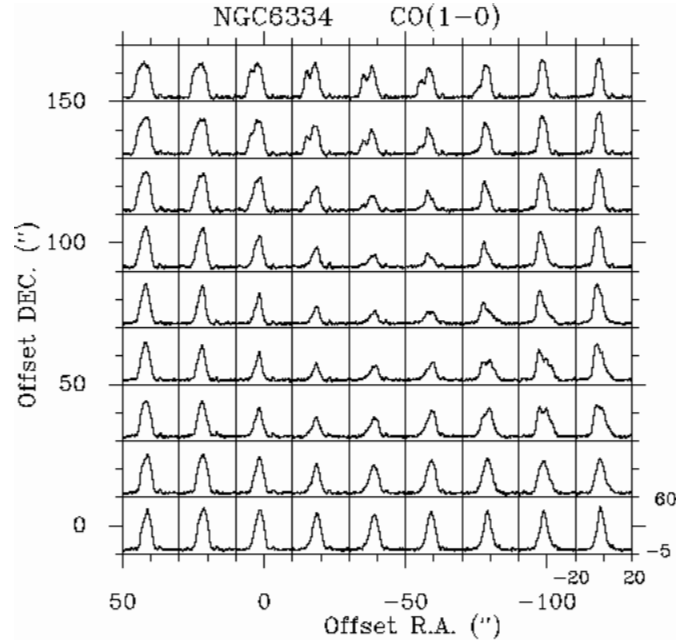}}}
\caption[]{\label{co10spec} CO $J$=1--0 spectra near the molecular void (for the full area, see Fig.\,\ref{spec}) showing red- and blueshifted emission
in addition to the main component. The displayed velocity range is --20\,km\,s$^{-1}$ $<$ $V_{\rm LSR}$ $<$ +20\,km\,s$^{-1}$. The
$T_{\rm mb}$ range is --5\,K $<$ $T_{\rm mb}$ $<$ +60\,K. The reference position is $\alpha_{1950} = 17^{\rm h} 17^{\rm m} 30\fs 0$,
$\delta_{1950} = -35^{\rm o} 47'30''$ ($\alpha_{2000}=17^{\rm h} 20^{\rm m} 51\fs 2$, $\delta_{2000}=-35^{\rm o} 50'28''$). The channel
spacing is 0.11\,km\,s$^{-1}$.}
\end{figure}

A detailed look at some CO spectra from the NGC\,6334 FIR II region, presented in Fig.\,\ref{co10spec}, reveals velocity components in
addition to the main feature discussed in Sect.\,3.1 and displayed in Fig.\,\ref{13co10v}. Beside this main component at --6.5 to
--2\,km\,s$^{-1}$, there is a weak +6\,km\,s$^{-1}$ feature that is only seen in CO and that is present in the N, E, S, and center.
Apparently, this component is related neither to the molecular void nor to the ridge of dense molecular gas in the southwestern part of
our map. In the north, there is a more notable component with a velocity of about --10\,km\,s$^{-1}$. It is best seen in the
$^{13}$CO($2-1$) line (Fig.\,\ref{spec}), but is also prominent in CO (Fig.\,\ref{co10spec}), HCN, and HCO$^+$, as well as in [C\,{\sc i}] near
the northern edge of the cross (Fig.\,\ref{CI}). This component marks the southern tip of a kinematic feature extending southwards from NGC\,6334
FIR\,I (see Fig.\,11 of Kraemer \& Jackson \cite{kraemer99}).

\begin{figure}
\resizebox{\hsize}{!}{\includegraphics{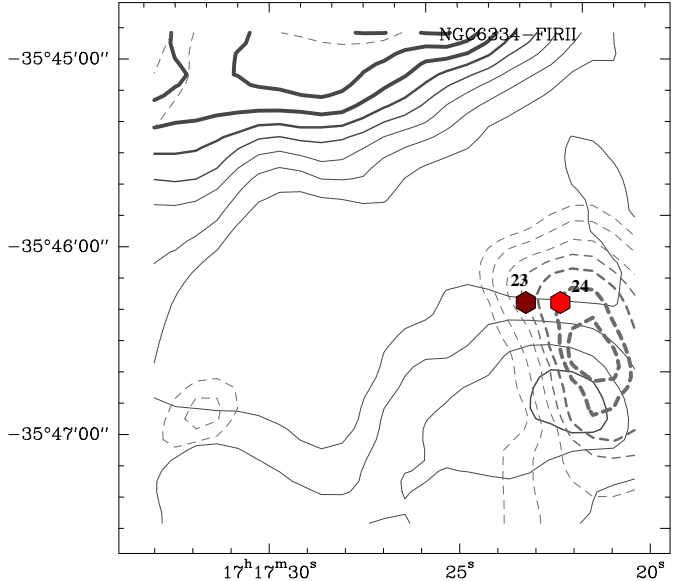}}
\caption[]{\label{outflow} Map of the red- and blueshifted CO components. Velocity ranges are --14\,km\,s$^{-1}$ $<$ $V_{\rm LSR}$ $<$
--8\,km\,s$^{-1}$ for the blue linewings (solid contours) and 2\,km\,s$^{-1}$ $<$ $V_{\rm LSR}$ $<$ 12\,km\,s$^{-1}$ for the red linewings
(dashed contours). Contour levels are 30\%--90\% of the integrated peak emission of 51\,K\,km\,s$^{-1}$ (blue) and 41\,K\,km\,s$^{-1}$ (red)
in steps of 10\%. The two hexagonal dots mark the stellar sources IRS23 and IRS24.}
\end{figure}

There also seems to be some evidence for a molecular outflow. Redshifted linewings of the `main' feature are apparent near the southwestern edge of
the molecular void, not only in CO but also in $^{13}$CO, HCN, and HCO$^+$. Figure~\ref{outflow} shows the CO map of this redshifted emission.
The redshifted linewing extends from the stellar sources IRS--23 and IRS--24 southwards, along the western edge of the molecular cloud
ridge (see Fig.\,\ref{cnt}) that shows high (column) densities (Sect.\,3.1). The corresponding blue-shifted linewings are, however, missing 
(Fig.\,\ref{outflow}). Furthermore, there is no indication of an outflow associated with the vibrationally excited H$_2$ gas observed by Straw 
\& Hyland (\cite{strawb}).

\subsection{$^{18}$O/$^{17}$O ratios}

The determination of CNO isotope ratios plays a key role in our understanding of stellar nucleosynthesis and chemical evolution (e.g., Wilson
\& Rood \cite{wilsonr}). One of the most interesting CNO isotope ratios is that of the two rare oxygen nuclei, $^{18}$O and $^{17}$O. With
$^{18}$O being mainly synthesized in massive stars ($\sim$8--20\,M$_{\odot}$) and $^{17}$O originating predominantly from lower mass stars
(e.g., Hoffman et al. \cite{hoffman}; Stoesz \& Herwig \cite{stoesz}), their abundance ratio may become a tracer of material ejected by high-mass 
stars.

Observing the $J$=1--0 lines of C$^{18}$O and C$^{17}$O has the advantage that both lines tend to be optically thin and that measured ratios are
small ($<$10) so that required observational sensitivities are not extremely different. Fractionation due to ion exchange reactions (see Watson
et al. \cite{watson} for the stable carbon isotopes) does not play a major role (Langer et al. \cite{langer}). Hence the integrated intensity
ratio is a good measure of $^{18}$O/$^{17}$O ratios (see Wouterloot et al. \cite{wouterloot} for an extensive multi-level analysis). While the
ratio is small (1.6$\pm$0.3) in the Large Magellanic Cloud, in the nuclear starbursts of NGC\,253 and NGC\,4945 $^{18}$O/$^{17}$O ratios of
$\sim$6.5 have been measured (Heikkil{\"a} et al. \cite{heikkilae}; Harrison et al. \cite{harrison}; Wang et al. \cite{wang}). These results
indicate that the ratio does not only depend on the relative importance of massive versus less massive stars, but that metallicity also plays
a role. In the Galaxy, the ratio is of order 4 (Table~\ref{compare}), but there is a clear discrepancy with the solar system (5.5; e.g., Anders
\& Grevesse \cite{anders}) that may have been formed out of material enriched by nearby massive stars (e.g., Lyons \& Young \cite{lyons}).

From our spectral data integrated intensities were obtained by Gaussian fitting or, in cases of poor signal-to-noise ratios, by the first moment
(area). In the latter cases we determined integrated intensities by summing line intensities over the velocity interval --13 to +3\,km\,s$^{-1}$.
C$^{18}$O/C$^{17}$O ratios were calculated for those 152 of the 169 positions that show C$^{17}$O signal-to-noise ratios $\ga$3.

\begin{figure}
\resizebox{\hsize}{!}{\includegraphics{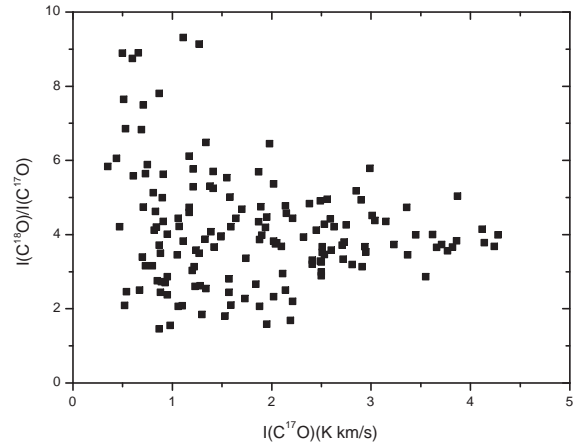}}
\caption[]{\label{ratioi} C$^{18}$O/C$^{17}$O integrated line intensity ratios versus C$^{17}$O $J$=1--0 integrated intensity.}
\label{ratio}
\end{figure}

\begin{table*}
\caption[]{Measured $^{18}$O/$^{17}$O ratios$^{\rm a)}$
\label{compare}}
\begin{flushleft}
\begin{tabular}{cccc}
\hline\noalign{\smallskip}
Source & Ratio & Notes & Reference \\
\hline\noalign{\smallskip}
Galactic disk                         & 3.65$\pm$0.15 & frequency corrected value          & Penzias (\cite{penzias})             \\
Nearby dark clouds                    & 3.2$\pm$0.2   & frequency corrected value          & Wilson et al. (\cite{wilsonlg})      \\
Ophiuchus cloud complex (1 position)  & 4.15$\pm$0.52 & frequency corrected value          & Bensch et al. (\cite{bensch})        \\
Taurus molecular cloud (648 positions)& 4.0$\pm$0.5   & from a ``simple model''            & Ladd (\cite{ladd})                   \\
Ophiuchus cloud complex (20 positions)& 3.53$\pm$0.11 & frequency corrected value          & Wouterloot et al. (\cite{wouterloot})\\
Ophiuchus cloud complex (17 positions)& 4.07$\pm$0.32 & assuming LTE                       & Wouterloot et al. (\cite{wouterloot})\\
Ophiuchus cloud complex (21 positions)& 4.10$\pm$0.14 & from an LVG model                  & Wouterloot et al. (\cite{wouterloot})\\
NGC\,6334\,FIR\,II (169 positions)    & 4.13$\pm$0.13 & frequency corrected value          & this paper                           \\

\noalign{\smallskip} \hline \noalign{\smallskip}
\end{tabular}
\end{flushleft}

a) Results are mainly derived from C$^{18}$O/C$^{17}$O $J$=1--0 integrated intensity ratios. Exceptions are Bensch et al. (\cite{bensch}), who
measured a $^{13}$C$^{18}$O/$^{13}$C$^{17}$O $J$=1--0 ratio, and Wouterloot et al. (\cite{wouterloot}), who analyzed the $J$=1--0, 2--1,
and 3--2 lines of C$^{18}$O and C$^{17}$O including some lines-of-sight with moderately optically thick C$^{18}$O emission.

\end{table*}

In Fig.\,\ref{ratio}, the ratios are plotted against the corresponding integrated intensities of C$^{17}$O. The ratios show significant scatter
at low integrated intensities due to noise in the C$^{17}$O spectra. For clouds filling the beam, total C$^{18}$O and C$^{17}$O column 
densities are proportional to $\nu^{-2}$ times the integrated line intensity (e.g., Linke et al. 1977). Measured line intensity ratios thus 
have to be corrected: C$^{18}$O/C$^{17}$O(corrected) = ($\nu_{\rm C17O}/\nu_{\rm C18O})^2$ $\times$ C$^{18}$O/C$^{17}$O(observed). The 
correction factor is 1.047. Applying this factor, we obtain abundance ratios of 4.32$\pm$0.13 averaging all plotted data points, 
4.13$\pm$0.13 for the positions with $I$(C$^{17}$O)$>$1\,K\,km\,s$^{-1}$, 4.02$\pm$0.10 for those with $I$(C$^{17}$O)$>$ 2\,K\,km\,s$^{-1}$, 
and 4.13$\pm$0.12 for those with $I$(C$^{17}$O)$>$3\,K\,km\,s$^{-1}$, the errors denoting the standard deviation of the mean. While the number 
of values obtained toward positions with increasing minimum intensity is decreasing, the scatter in the individual ratios is also decreasing 
because of higher signal-to-noise ratios. Thus standard deviations remain almost the same.

The derived values are in excellent agreement with those of the LTE and LVG (large velocity gradient) model results presented by Wouterloot et
al. (\cite{wouterloot}) from the Ophiuchus cloud, with that reported by Bensch et al. ({\cite{bensch}) from a $^{13}$C$^{18}$O/$^{13}$C$^{17}$O 
intensity ratio toward the same cloud complex (see Table~\ref{compare}), and with those of Ladd et al. (\cite{ladd}) from the similarly nearby 
Taurus molecular cloud. Our frequency-corrected values are higher than those of the ``only'' frequency-corrected values of Wouterloot et al.
(\cite{wouterloot}), but this is not a serious disagreement. Wouterloot et al. (\cite{wouterloot}) purposely observed positions with a wide range
of column densities, including some positions where the C$^{18}$O opacity was not negligible. In these cases, the linear correlation between line
strength and column density breaks down and abundance ratios become larger than frequency-corrected line intensity ratios. This effect {\it may}
also play a role in the data presented by Penzias (\cite{penzias}) and Wilson et al. (\cite{wilsonlg}).

Our data do not hint at optical depth effects. The C$^{18}$O/C$^{17}$O ratio stays constant up to the positions with highest line intensity. 
$^{18}$O enriched material from massive stars and isotope selective photodissociation would increase the measured line intensity ratios. High 
opacities in C$^{18}$O would reduce them. Only having measured the $J$=1--0 lines of C$^{18}$O and C$^{17}$O, we cannot firmly exclude that 
these three effects precisely cancel each other. However, in view of our estimate of $\tau$($^{13}$CO $J$=1--0) $\sim$ 0.3 in Sect.\,4.4 and 
following the principle of Occam's razor, such a scenario is farfetched. Line saturation and isotope selective photodissociation can thus be
neglected.

The aim of the isotope ratio measurements was to find out, whether $^{18}$O enrichment can be seen at the very edge of the molecular void. This is
neither apparent in Fig.\,\ref{ratio} nor is there any such tendency when only analyzing those positions that are located next to the H\,{\sc ii}
region. At the eastern edge of the void, averaging spectra at positions with offsets (0,40), (0,60), (0,80), and (0,100) (see Figs.\,\ref{outflow}
and \ref{spec}), we find a frequency-corrected ratio of 3.0$\pm$0.2. Toward its western edge, at positions (--80,40), (--80,60), (--80,80),
(--80,100), and (--80,120), we find 2.3$\pm$0.1. The latter ratio appears to be too low because toward one position, C$^{17}$O but not C$^{18}$O,
is detected. Thus pointing uncertainties {\it might} play a role. The ring with radius 40$''$--50$''$ and density $n$(H$_2$) $\sim$ 10$^3$\,cm$^{-3}$ 
around the center of the void (representing the inner edge of the cloud, see Sects.\,3 and 4.4) contains about 5--10\,M$_{\odot}$, which is similar 
to the mass formerly filling the ionized bubble, if we assume the same average density. Depending on stellar mass and evolutionary stage, the 
gas at the edge of the void {\it might} therefore be significantly affected by the ejecta (possibly also amounting to several M$_{\odot}$) of young massive 
stars. We note, however, that the $^{18}$O/$^{17}$O ratios near the molecular void suggest the presence of {\it lower}, not higher $^{18}$O/$^{17}$O 
ratios when comparing them with values from interstellar clouds in the solar neighborhood. This does not agree with the high values known to 
characterize extragalactic nuclear starbursts}.

\subsection{Formation and evolutionary stage of the molecular void}

As already indicated, in nuclear starbursts of nearby galaxies $^{18}$O is enriched with respect to $^{17}$O (Harrison et al. \cite{harrison};
Wang et al. \cite{wang}). In the Orion OB association the youngest sub-associations show higher oxygen abundances than the older sub-associations
(Cunha \& Lambert \cite{cunha}). The solar system shows a significantly enhanced $^{18}$O/$^{17}$O ratio relative to interstellar values, presumably
because of $^{18}$O enrichment just prior to its formation (Henkel \& Mauersberger \cite{henkel}). Why do we not measure enhanced $^{18}$O/$^{17}$O
ratios near the edge of the NGC\,6334-FIRII molecular cloud? Following Heger \& Langer (\cite{heger}), Hoffman et al. (\cite{hoffman}), 
and Rauscher et al. (\cite{rauscher}), stars with zero age main sequence masses up to 20\,M$_{\odot}$ show high $^{18}$O yields during core collapse. 
With a highly obscured O6.5 star (IRS--24; $M$$\sim$40\,M$_{\odot}$) likely being responsible for the molecular void (Sect.\,4.1), stars with 
$\la$20\,M$_{\odot}$ had, however, not yet sufficient time to reach the supernova stage of evolution. For rotating massive stars, Heger \& Langer 
(\cite{heger}) predict enhanced $^{17}$O and depleted $^{18}$O surface abundances during the pre-supernova stages, at the time of the first dredge 
up and even before, at core hydrogen exhaustion. Prior to core collapse, a low $^{18}$O/$^{17}$O ratio in the affected interstellar medium is 
therefore consistent with theory.

The lifetime of an O6.5 star is $\sim$4.5$\times$10$^6$\,yr (e.g., Massey \cite{massey}), providing an upper age limit for the molecular void. There
are also indications that the region is not extremely young (Sect.\,4.1). The extent of the FIR emission and the size of the radio continuum source
($\lambda$16\,cm diameter: $\sim$1\,pc; e.g., Brooks \& Whiteoak \cite{brooks}) are large when compared with the other main sources of activity in
NGC\,6334. Remarkable is the regular structure of the radio continuum shell. Apparently, while being very clumpy on small scales (Sect.\,4.5), the
surrounding parental molecular cloud must have been quite homogeneous on larger scales. There is no hint for the presence of a ``champagne phase''
expected when the ionized shell expands to the edge of the cloud.

As already indicated in Sect.\,1, H\,{\sc ii} regions expanding into an approximately homogeneous medium should show, at later stages of their evolution,
a decelerated subsonic ``D-type'' ionization front that is slowly following a shock front farther out. This is likely what we are viewing. While thermal 
continuum radiation encompasses an almost circular area of diameter 2\arcmin, even further to the east there is the prominent arc of warm presumably
shocked H$_2$ gas. Since CO is not providing any evidence for outflowing material, H$_2$ may not trace the location of an outflow (as suggested by 
Straw \& Hyland \cite{strawb}), but instead marks the shockfront compressing neutral gas at the periphery of the evolved H\,{\sc ii} region 
surrounding IRS--24.

In Sects.\,4.4 and 4.5 we obtained cloud densities of 10$^{3}$\,cm$^{-3}$ from CO and $\ga$10$^{5}$\,cm$^{-3}$ from HCN. For an intermediate value, 
$n$(H$_2$) = 10$^4$\,cm$^{-3}$, Fig.\,\ref{finder}c provides the size of the ``initial'' Str{\"o}mgren sphere, 0.35\,pc or $0\,.\!\!^{\prime}$7. This 
is smaller than the observed sizes of the molecular void and H\,{\sc ii} region, $\sim$$1\,.\!\!^{\prime}$5 and 2\arcmin\ (Figs.\,\ref{cnt} and 
\ref{finder}). Since an H\,{\sc ii} region can expand up to about five times its Str{\"o}mgren radius (until pressure equilibrium is attained) if 
the ionizing star does not move off the main sequence before (e.g., Yorke 1986), there is no serious contradiction between models and observations. 
In case the ionized gas just occupies the ``initial'' Str{\"o}mgren sphere, we obtain an average density of $n$(H$_2$) $\sim$ 2$\times$10$^{3}$\,cm$^{-3}$ 
(e.g., Tielens 2005, p.\,231). This is consistent with our density estimate from CO, which represents the bulk of the molecular gas surrounding the void.

In between the outer shell of shock heated vibrationally excited H$_2$ (Straw \& Hyland \cite{strawb}) and the ionized region further inside,
conditions should be optimal for PDRs. While [C\,{\sc i}] is weak in the molecular void that is readily observed in the radio continuum through
thermal free-free emission, [C\,{\sc ii}] is quite prominent (Kraemer et al. \cite{kraemer00}). Suitable discriminators between PDRs and shocks
are the absolute and relative intensities of the [C\,{\sc ii}] 158\,$\mu$m and [O\,{\sc i}] 63\,$\mu$m fine structure lines. One can find weak
[C\,{\sc ii}] and strong [O\,{\sc i}] emission ($I$([O\,{\sc i}])/$I$([C\,{\sc ii}])$\gg$10) in shocks, while the reverse holds for PDRs (Genzel
\cite{genzel}). Observations of far-infrared fine-structure lines in NGC\,6334, [C\,{\sc ii}] 158\,$\mu$m, and [O\,{\sc i}] 63\,$\mu$m and 145\,$\mu$m,
strongly support the presence of PDRs that must be widespread. The most intense [C\,{\sc ii}] emission in NGC\,6334 is associated with the FIR\,II
region (Boreiko \& Betz \cite{boreiko}; Kraemer et al. \cite{kraemer98}, \cite{kraemer00}). $I$([O\,{\sc i}])/$I$([C\,{\sc ii}]) $\sim$ 3 (Kraemer et
al. 1998).

\section{Summary}

Having performed observations of molecular and atomic lines in the galactic star-forming region NGC\,6334, we find the following main results:

(1) The large-scale distribution of $^{13}$CO $J$=1--0 emission over the entire star forming-region shows strong peaks toward FIR\,I, IV, and V,
while FIR\,II and III are merely associated with ridge emission extending southwards from FIR\,I and northeastwards from FIR\,IV, respectively.

(2) A detailed study of the eastern part of the most luminous source of far-infrared emission, NGC\,6334 FIR\,II, reveals a molecular void
associated with the H\,{\sc ii} region NGC\,6334--D. Molecular and [C\,{\sc i}] excitation indicate $T_{\rm kin}$ $\sim$40--50\,K in the surrounding 
molecular cloud, while toward the void itself $T_{\rm kin}$ $\ga$ 200\,K. Molecular column densities and measured obscuration suggest that the
exciting star IRS--24 (presumably of type O6.5) is located behind the bulk of the molecular column.

(3) The CO spectra show redshifted velocity wings south of IRS--24, but the corresponding blueshifted feature is missing.

(4) $^{18}$O/$^{17}$O abundance ratios reflecting the $^{18}$O/$^{17}$O isotope ratio were obtained from measurements of the $J$=1--0 
lines of C$^{18}$O and C$^{17}$O. The mean ratio is 4.13$\pm$0.13, in good agreement with recent measurements toward nearby molecular clouds. At 
the edge of the molecular void, however, ratios might be smaller.

(5) The regular morphology of the ionized gas of the H\,{\sc ii} region indicates the presence of a rather homogeneous parental molecular cloud on 
parsec scales. The D-type ionization front (from observations of the radio continuum) and presumably also the shock front farther out in the 
molecular gas (from measurements of vibrationally excited H$_2$) can be identified. The age of the H\,{\sc ii} region must be $<$5$\times$10$^6$\,yr.

(6) The [C\,{\sc i}] distribution follows that of the molecular gas and does not show a bright rim at the inner portion of the molecular shell. 
This may be interpreted in terms of a high degree of small-scale clumping, with PDRs reaching deep into the dense surrounding cloud. Such small 
scale clumping is indicated by the different densities obtained from CO and HCN data ($n$(H$_2$)$\sim$10$^3$ and $\ga$10$^5$\,cm$^{-3}$, 
respectively). Observationally, however, the [C\,{\sc i}] lines do not provide any direct hint for a close connection with PDRs.

So far, the velocity field of the shocked H$_2$ gas has not been measured. Additional observations of tracers of shocked molecular gas like SiO 
(Gueth et al. \cite{gueth}) or HNCO (Zinchenko et al. \cite{zinchenko}) would also be desirable. A dedicated experiment, investigating 
the potential anomaly in the $^{18}$O/$^{17}$O ratio near the edge of the void to further constrain enrichment by ejecta from massive stars would 
be another worthwhile project.

\begin{acknowledgements}
We wish to thank the anonymous referee for useful comments. J.S. Zhang acknowledges support by the exchange program between the Chinese Academy of 
Sciences and the Max-Planck-Gesellschaft.
\end{acknowledgements}

\Online
\appendix
\onecolumn

\section{Observed SEST spectra}

\begin{figure*}
\resizebox{15cm}{22cm}{\includegraphics{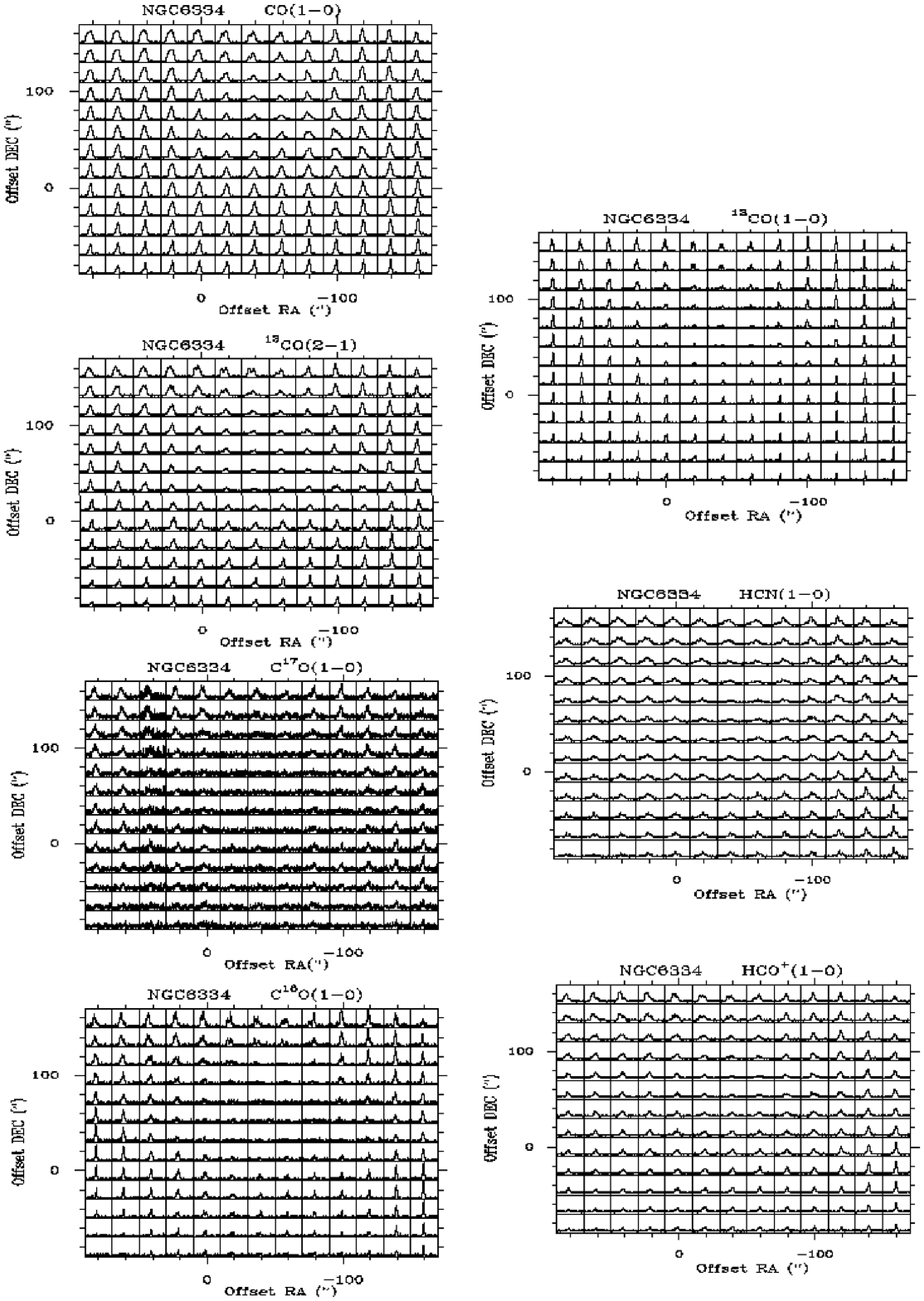}}
\caption{Spectra of seven molecular transitions across the NGC\,6334 FIR II region. Left (from top to bottom): $^{12}$CO(1-0), $^{13}$CO(2-1),
C$^{17}$O,  and  C$^{18}$O; Right: $^{13}$CO(1-0), HCN(1-0), and HCO$^{+}$(1-0). The (0,0) position of the maps is $\alpha_{1950}=17^{\rm h}
17^{\rm m} 30\fs 0$ and $\delta_{1950}=-35^{\rm o} 47'30''$ ($\alpha_{2000}=17^{\rm h} 20^{\rm m} 51\fs 2$ and $\delta_{2000}=-35^{\rm o} 50'28''$).
The x-axis of each individual spectrum denotes the local standard of rest velocity from --20 to +20\,km\,s$^{-1}$. Main beam brightness temperature
scales are --5 to 60\,K (CO), --1 to 25\,K ($^{13}$CO $J$=1--0), --5 to 45\,K ($^{13}$CO $J$=2--1), --0.2 to 1\,K (C$^{17}$O), --0.2 to 4\,K
(C$^{18}$O), and --2 to 13\,K (HCN and HCO$^+$).}
\label{spec}
\end{figure*}


\begin{thebibliography}{}

\bibitem[1989]{anders} Anders, E., \& Grevesse, N. 1989, Geochim. Cosmochim. Acta, 53, 197
\bibitem[2001]{bensch} Bensch, F., Pak, I., Wouterloot, J. G. A., Klapper, G., \& Winnewisser, G. 2001, ApJ, 562, L185
\bibitem[1978]{bohlin} Bohlin, R. C., Savage, B. D., \& Drake, J. F. 1978, ApJ, 224, 132
\bibitem[1989]{booth} Booth, R. S., Delgado, G., \& Hagstr\"om, M., et al. 1989, A\&A, 216, 315
\bibitem[1995]{boreiko} Boreiko, R. T., \& Betz, A. L., 1995, ApJ, 454, 307
\bibitem[2001]{brooks} Brooks, K. J., \& Whiteoak, J. B. 2001, MNRAS, 320, 465
\bibitem[2005]{brouillet} Brouillet, N., Muller, S., Herpin, F., Braine, J., \& Jacq, T. 2005, A\&A, 429, 153
\bibitem[2000]{burton} Burton, M. G., Ashley, R. D., \& Marks, R. D. et al. 2000, ApJ, 542, 359
\bibitem[2006]{bykov} Bykov, A. M., Krassilchtchikov, A. M., Uvarov, Y.A., et al. 2006, A\&A, 449, 917
\bibitem[2000]{caproni} Caproni, A., Abraham, Z., \& Vilas-Boas, J. W. S. 2000, A\&A, 361, 685
\bibitem[2002]{carral} Carral, P., Kurtz, S. E., Rodr\'{\i}guez, L. F., et al. 2002, ApJ, 123, 2574
\bibitem[1997]{chin} Chin, Y.-N., Henkel, C., Whiteoak, J. B., et al. 1997, A\&A, 317, 548
\bibitem[2002]{church} Churchwell, E. 2002, ARA\&A, 40, 27
\bibitem[2005]{christopher} Christopher, M. H., Scoville, N. Z., Stolovy, S. R., \& Yun, M. S. 2005, ApJ, 622, 346
\bibitem[1992]{cunha} Cunha, K. \& Lambert, D. L. 1992, ApJ, 399, 586
\bibitem[1989]{downes} Downes, D., 1989, in: Introductory Course in galaxies' evolution and observational astronomy, eds.:
                       I. Appenzeller, H. Habing, P. Lena, Springer-Verlag, Heidelberg, p. 353
\bibitem[2006]{ezoe} Ezoe, Y., Kokubin, M., Makishima, K., Sekimoto, Y., \& Matsuzaki, K. 2006, ApJ 638, 860
\bibitem[2001]{flower} Flower, D. R. 2001, J. Phys. B: At. Mol. Opt. Phys., 34, 1
\bibitem[1990]{franco} Franco, J., Tenorio-Tagle, G., \& Bodenheimer, P. 1990, ApJ, 349, 126
\bibitem[1993]{fuente} Fuente, A., Mart\'{\i}n-Pintado, J., Cernicharo, J., \& Bachiller, R. 1993, A\&A, 276, 473
\bibitem[1982]{gezari} Gezari, D. 1982, ApJ, 259, L29
\bibitem[1990]{gezarib} Gezari, D., \& Blitz, L. 1990, in: Submillimetre Astronomy, eds: G.D.Watt, A.S.Webster, Kluwer, p.173
\bibitem[1991]{genzel} Genzel, R. 1991, in: The physics of star formation and early stellar evolution, eds. C.J. Lada \& N.D. Kylafis, p.155
\bibitem[1998]{gueth} Gueth, F., Guilloteau, S., \& Bachiller, R. 1998, A\&A, 333, 287
\bibitem[2000]{guillot} Guilloteau, S., \& Lucas, R., 2000, in Imaging at Radio through Submillimeter Wavelengths, ASP Conf. Ser., 217, 299
\bibitem[2006]{guesten} G{\"u}sten, R., Nyman, L. \AA., Schilke, P., et al.  2006, A\&A, 454, L13
\bibitem[1999]{harrison} Harrison, A., Henkel, C., \& Russell, A. 1999, MNRAS, 303, 157
\bibitem[2000]{heger} Heger, A., \& Langer, N. 2000, ApJ, 544, 1016
\bibitem[1998]{heikkilae} Heikkil{\"a}, A., Johansson, L. E. B., \& Olofsson, H. 1998, A\&A, 332, 493
\bibitem[1997]{helfer} Helfer, T. T., \& Blitz, L. 1997, ApJ, 478, 233
\bibitem[1993]{henkel} Henkel, C., \& Mauersberger, R. 1993, A\&A, 274, 730
\bibitem[2006]{heymin} Heyminck, S., Kasemann, C., G{\"u}sten, R., de Lange, G., \& Graf, U. U. 2006, A\&A, 454, L21
\bibitem[2001]{hoffman} Hoffman, R. D., Woosley, S. E., \& Weaver, T. A. 2001, ApJ, 549, 1085
\bibitem[1997]{hollen} Hollenbach, D. J., \& Tielens, A. G. G. M. 1997, ARA\&A, 35, 179
\bibitem[2006]{hunter} Hunter, T. R., Brogan, C. L., Megeath, S. T., Menten, K. M., \& Beuther, S. 2006, ApJ, 649, 888
\bibitem[2002]{ikeda} Ikeda, M., Oka, T., Tatematsu, K., Sekimoto, Y., \& Yamamoto, S. 2002, ApJS, 139, 467
\bibitem[2006]{klein} Klein, B., Philipp, S. D., Kr{\"a}mer, I., et al. 2006, A\&A, 454, L29
\bibitem[1998]{kraemer98} Kraemer, K. E., Jackson, J. M., \& Lane, A.P. 1998, ApJ, 503, 785
\bibitem[1999]{kraemer99} Kraemer, K. E., \& Jackson, J. M. 1999, ApJS, 124, 439
\bibitem[2000]{kraemer00} Kraemer, K. E., Jackson, J. M., Lane, A. P., \& Paglione, T. A. D. 2000, ApJ, 542, 946
\bibitem[1995]{kuiper} Kuiper, T. B. H., Peters\,III, W. L., Foster, J. R., Gardner, F. F., \& Whiteoak, J. B. 1995, ApJ, 446, 692
\bibitem[2004]{ladd} Ladd, E.F. 2004, ApJ, 610, 320
\bibitem[1984]{langer} Langer, W. D., Graedel, T. E., Frerking, M. A., \& Armentrout, P. B. 1984, ApJ, 277, 581
\bibitem[2006]{leurini} Leurini, S., Schilke, P., Parise. F., et al. 2006, A\&A, 454, L83
\bibitem[1977]{linke} Linke, R. A., Goldsmith, P. F., Wannier, P. G., Wilson, R. W. \& Penzias, A. A. 1977, ApJ, 214, 50
\bibitem[1986]{loughran} Loughran. L., McBreen, B., Fazio, G. G. et al., 1986, ApJ, 303, 629
\bibitem[2005]{lyons} Lyons, J. R., \& Young, E. D. 2005, Nat, 435, 317
\bibitem[2000]{mao} Mao, R. Q., Henkel, C., Schulz, A., et al. 2000, A\&A, 358, 433
\bibitem[2005]{massey} Massey, P. 2005, ARA\&A, 41, 15
\bibitem[1979]{mcbreen} McBreen, B., Fazio, G. G., Stier, M., \& Wright, E. L. 1979, ApJ, 232, L183
\bibitem[2000]{mccutcheon} McCutcheon, W. H., Sandell, G., Matthews, H. E., et al. 2000, MNRAS, 316, 152
\bibitem[1982]{meaburn} Meaburn, J., \& White, N. J. 1982, ApJ, 255, L55
\bibitem[1980]{moran}  Moran, J.M., \& Rodr\'{\i}guez, L.F. 1980, ApJ, 236, L159
\bibitem[1978]{neckel} Neckel, T. 1978, A\&A, 69, 51
\bibitem[1981]{penzias} Penzias, A. A. 1981, ApJ, 249, 518
\bibitem[2000]{persi} Persi, P., Tapia, M., \& Roth, M. 2000, A\&A 357, 1020
\bibitem[1999]{pirogov} Pirogov, L. 1999, A\&A, 348, 600
\bibitem[1999]{plume} Plume, R., Jaffe, D. T., Tatematsu, K., Evans\,II, N. J., \& Keene, J. 1999, ApJ, 512, 768
\bibitem[2002]{rauscher} Rauscher, T., Heger, A., Hoffman, R. D., \& Woosley, S. E. 2002, ApJ, 576, 323
\bibitem[1982]{rodriguez} Rodr\'{\i}guez, L., Cant\'o, J., \& Moran, J. 1982, ApJ, 255, 103
\bibitem[1996]{rohlfs} Rohlfs, K., \& Wilson, T. L. 1996, Tools of Radio Astronomy, 2nd edition, Springer, Berlin, p194
\bibitem[2000]{sandell} Sandell, G. 2000, A\&A, 358, 242
\bibitem[2000]{sarma} Sarma, A. P., Troland, T. H., Roberts, D. A., \& Crutcher, R. M. 2000, ApJ, 533, 271
\bibitem[2006]{schilke} Schilke, P., Comito, C., Thorwirth, S., et al. 2006, A\&A, 454, L41
\bibitem[2003]{schneider} Schneider, N., Simon, R., Kramer, C. et al. 2003, A\&A, 406, 915
\bibitem[2005]{schoier} Sch{\"o}ier, F. L., van der Tak, F. F. S., van Dishoeck, E. F., Black, J. H. 2005, A\&A, 432, 369
\bibitem[1993]{handbook} SEST Handbook, 1993, European Southern Observatory Operating Manual No. 19, Version~1.0
\bibitem[2003]{stoesz} Stoesz, J. A., \& Herwig, F. 2003, MNRAS, 340, 763
\bibitem[1989a]{strawa} Straw, S. M., \& Hyland, A. R. 1989a, ApJ, 340, 318
\bibitem[1989b]{strawb} Straw, S. M., \& Hyland, A. R. 1989b, ApJ, 342, 876
\bibitem[1989c]{strawc} Straw, S. M., Hyland, A. R., \& McGregor, P. J. 1989c, ApJS, 69, 99
\bibitem[1997]{stutzki} Stutzki, J., Graf, U. U., Haas, S. et al. 1997, ApJ, 477, L33
\bibitem[1988]{tenorio} Tenorio-Tagle, G. 1988, ARA\&A, 26, 145
\bibitem[2005]{Tiel05} Tielens, A. G. G. M. 2005, The physics and chemistry of the interstellar medium, Cambridge, Cambridge University Press
\bibitem[2004]{wang} Wang, M., Henkel, C., Chin, Y.-N., et al. 2004, A\&A, 422, 883
\bibitem[1976]{watson} Watson, W. D., Anicich, V. G., \& Huntress, W. T. 1976, ApJ, 205, L165
\bibitem[2003]{weiss} Wei{\ss}, A., Henkel, C., Downes, D., \& Walter, F. 2003, A\&A, 409, L41
\bibitem[1981]{wilsonlg} Wilson, R. W., Langer,W. D., \& Goldsmith, P. F. 1981, ApJ, 243, L47
\bibitem[1994]{wilsonr} Wilson, T. L., \& Rood, R. T. 1994, ARA\&A, 32, 191
\bibitem[1989]{wilsonw} Wilson, T. L., \& Walmsley, C. M. 1989, A\&AR, 1, 141
\bibitem[2005]{wouterloot} Wouterloot, J.G.A., Brand, J., \& Henkel, C. 2005, A\&A, 430, 549
\bibitem[1986]{Yorke} Yorke, H. W. 1986, ARA\&A, 24, 49
\bibitem[2000]{zinchenko} Zinchenko, I., Henkel, C., \& Mao, R. Q. 2000, A\&A, 361, 1079

\end{thebibliography}
\end{document}